\documentclass[12pt,twocolumn]{aastex63}
\newcommand{\kms}{{\rm km~s}$^{-1}$}
\newcommand{\kmsMpc}{{km s$^{-1}$ Mpc$^{-1}$}}

\newcommand{\dn}{D_{n}4000}

\usepackage{color}
\usepackage{multirow}
\usepackage{graphicx}
\usepackage{amsmath}
\usepackage{hhline}

\begin{document}

\title{Tracing Dark Matter Halos with Satellite Kinematics and the Central Stellar Velocity Dispersion of Galaxies}

\author{Gangil Seo$^{1}$,
             Jubee Sohn$^{2}$, 
             Myung Gyooon Lee$^{1, *}$}

\affil{$^{1}$ Astronomy Program, Department of Physics and Astronomy, Seoul National University, Gwanak-gu, Seoul 151-742, Republic of Korea}
\affil{$^{2}$ Smithsonian Astrophysical Observatory, 60 Garden Street, Cambridge, MA 02138, USA}

\email{$^{*}$corresponding author : mglee@astro.snu.ac.kr}

\begin{abstract}
It has been suggested that the central stellar velocity dispersion of galaxies 
 can trace dark matter halo mass directly. 
We test this hypothesis using a complete spectroscopic sample of 
 isolated galaxies surrounded by faint satellite galaxies 
 from the Sloan Digital Sky Survey Data Release 12. 
We apply a friends-of-friends algorithm with projected linking length $\Delta D < 100$ kpc and 
 radial velocity linking length $\Delta V < 1000$ km s$^{-1}$~ to construct our sample. 
Our sample includes 2807 isolated galaxies with 3417 satellite galaxies at $0.01 < z < 0.14$. 
We divide the sample into two groups based on the primary galaxy color:
 red and blue primary galaxies separated at $(g-r)_{0} = 0.85$. 
The central stellar velocity dispersions of the primary galaxies are proportional to 
 the luminosities and stellar masses of the same galaxies. 
Stacking the sample based on the central velocity dispersion of the primary galaxies, 
 we derive the velocity dispersions of their satellite galaxies,
 which trace the dark matter halo mass of the primary galaxies. 
The system velocity dispersion of the satellite galaxies shows a remarkably tight correlation 
 with the central velocity dispersion of the primary galaxies for both red and blue samples. 
In particular, the slope of the relation is identical to 1 for red primary systems. 
This tight relation suggests that the central stellar velocity dispersion of galaxies 
 is indeed an efficient and robust tracer for dark matter halo mass. 
We provide empirical relations between the central stellar velocity dispersion and the dark matter halo mass.
\end{abstract}

\section{Introduction}

\subsection{Tracing Dark Matter Halo Mass}
According to the modern galaxy formation scenario based on $\Lambda$CDM cosmology, 
 galaxies are formed and embedded in dark matter halos, 
 so there must be close relations between galaxy properties and dark matter halo properties 
 (\citealp{wec18,beh19,bur20} and references therein). 
However, it is not easy to derive directly the extent and total mass of dark matter around galaxies.

Various methods have been applied to estimate the amount of dark matter in galaxies and
 groups/clusters of galaxies (see reviews like \citealp{cou14,sal19} and references therein).
Traditionally, dynamical tracers 
 (stars, globular clusters, planetary nebulae, HII regions, HI gas, or satellite galaxies) 
 have been used for mass estimation. 
In particular, the kinematics of satellite galaxies among them has been a powerful tool to study
 the dark matter in the Local Group \citep{lyn81, wat10, mcc12}. 
However, this method can be applied only to a small number of nearby galaxies,
 and they must have a reasonable number of satellite galaxies \citep{wat10, wec18}. 

\subsection{Satellite Kinematics}
A number of previous studies used statistical (or stacking) analysis of satellite kinematics for 
 a large number of isolated galaxies in order to investigate the property of dark matter halos \citep{zar93,zar97,mck02,pra03,bra05,con07,kly09,dut10,mor11,woj13,van16,lan19}.
However, this approach can only be applied to galaxies with satellites, which are only found in the local universe.

Therefore, simple proxies that can be applied to a large sample of galaxies are often used to estimate dark matter halo mass indirectly. It is known that the luminosity or stellar mass of galaxies shows a correlation with the dark matter halo mass derived from their satellite observations, but their relations have been controversial \citep{nor08, woj13,van16}.

\subsection{Central Velocity Dispersion as a Proxy for Dark Matter Halo Mass}

In their pioneering paper, \citet{fab76} found that the central stellar velocity dispersion obtained from 25 elliptical and S0 galaxies increases with the galaxy $B$-band luminosity according to $L_{B} \propto \sigma_{0}^4$, the Faber-Jackson relation.
Later, \citet{whi79} found that the bulges of spiral galaxies follow a similar relation.
These studies and numerous following ones showed that the kinematic properties of the central region of a galaxy are linked with the properties of the entire galaxy (like luminosity and stellar mass; see \citet{dav19} and references therein).

Recently, it has been suggested that the central stellar velocity dispersion of a galaxy 
 can also be a robust proxy of the dark matter halo velocity dispersion and the dark matter halo mass 
 \citep{sch15, zah16, zah18, uts20}. 
Based on the Illustris-1 hydrodynamical cosmological simulations \citep{vog14}, 
 \citet{zah18} showed that the stellar velocity dispersion of quiescent  galaxies is proportional 
 to the dark matter halo velocity dispersion. 
The tight relation ($\sim$0.2 dex scatter) indicates that the dark matter halo mass can be estimated well 
 from the central stellar velocity dispersion of central (primary) galaxies.  
 
Several observational studies also suggest that the central stellar velocity dispersion is 
 a good tracer of dark matter halo mass \citep{wak12, zah16, soh20}. 
The central velocity dispersion has several advantages in estimating the dark matter halo mass
 over luminosity and stellar mass (e.g., \citealp{soh17a}). 
The central velocity dispersion is relatively insensitive to the 
systematic uncertainties \citep{soh17a}
 that can be introduced by photometry and the underlying physical models to derive photometry and stellar mass 
 (e.g., \citealp{con07, ber13}). 
\citet{zah16} demonstrated that the stellar mass and central stellar velocity dispersion relation has 
 a similar slope to that of the relation between the dark matter halo mass and dark matter velocity dispersion. 
This consistency suggests that the central stellar velocity dispersion is directly proportional to the dark matter halo velocity dispersion.
More recently, \citet{soh20} showed that 
 the central stellar velocity dispersion of the Brightest Cluster Galaxies (BCGs) 
 is tightly correlated with the system velocity dispersion of cluster galaxies. 
This tight relation supports the idea that the central stellar velocity dispersion of the central galaxy 
 can trace the dark matter halo mass.

\subsection{Goals of This Study}

In this study, 
 we present an empirical test of the above hypothesis that 
 the central stellar velocity dispersion of a galaxy 
 can be a robust proxy of the dark matter halo velocity dispersion and the dark matter halo mass, 
 using a large sample of isolated galaxies with satellite galaxies. 
We construct a complete sample of isolated galaxies surrounded by faint satellite galaxies 
 from the Sloan Digital Sky Survey (SDSS) Data Release 12 (DR12) spectroscopic data. 
Based on this complete sample, 
 we explore the relation between the satellite velocity dispersion 
 and the central stellar velocity dispersion of primary galaxies. 
We also study the dependence of the satellite velocity dispersion on other physical properties of their primary galaxies 
 (i.e., luminosity and stellar mass). 
We then derive the empirical relations between the central stellar velocity dispersion of the primary galaxy and 
 its dark matter halo mass.

This paper is organized as follows. 
In Section 2 we describe how we select samples of isolated galaxies with satellites from the SDSS data.
In \S3, we present the properties of the primary galaxies and 
 estimate the system velocity dispersion of satellite galaxies. 
Then, we investigate the relations between the satellite velocity dispersion and 
 the properties of the primary galaxies (central stellar velocity dispersion, stellar mass, luminosity, and dark matter halo mass) in \S4.
In \S5, we discuss the implications of the main results, 
 and we summarize the main results in \S6.
Throughout the paper, we adopt a standard $\Lambda$CDM cosmology 
 ($\Omega_{m} = 0.3$ and $\Omega_{\Lambda} = 0.7$) with $H_{0}=70$ \kmsMpc. 

\section{Data and Sample Selection}

\subsection{Data}

We used a catalog of galaxies at $0.01 \leq z < 0.14$ 
 based on the enhanced SDSS DR12 spectroscopic catalog, 
 which was used in the study of compact groups of galaxies by \citet{soh16}. 
A detailed description of the catalog is given in \citet{soh16, soh17a}. 
This catalog includes spectroscopic redshifts obtained from 
 the SDSS and the NASA/IPAC Extragalactic Database. 
For $r$-band absolute magnitudes and $(g-r)_0$ colors of galaxies,
 we use composite model (cModel) magnitudes.
Absolute magnitudes of galaxies are corrected for $k_{z=0.0}$ correction, evolution, 
 and foreground extinction.
We derive the stellar masses of galaxies using the LePHARE spectral energy distribution fitting code \citep{arn99,ilb06}.

\begin{figure}
\centering
\includegraphics[scale=0.32]{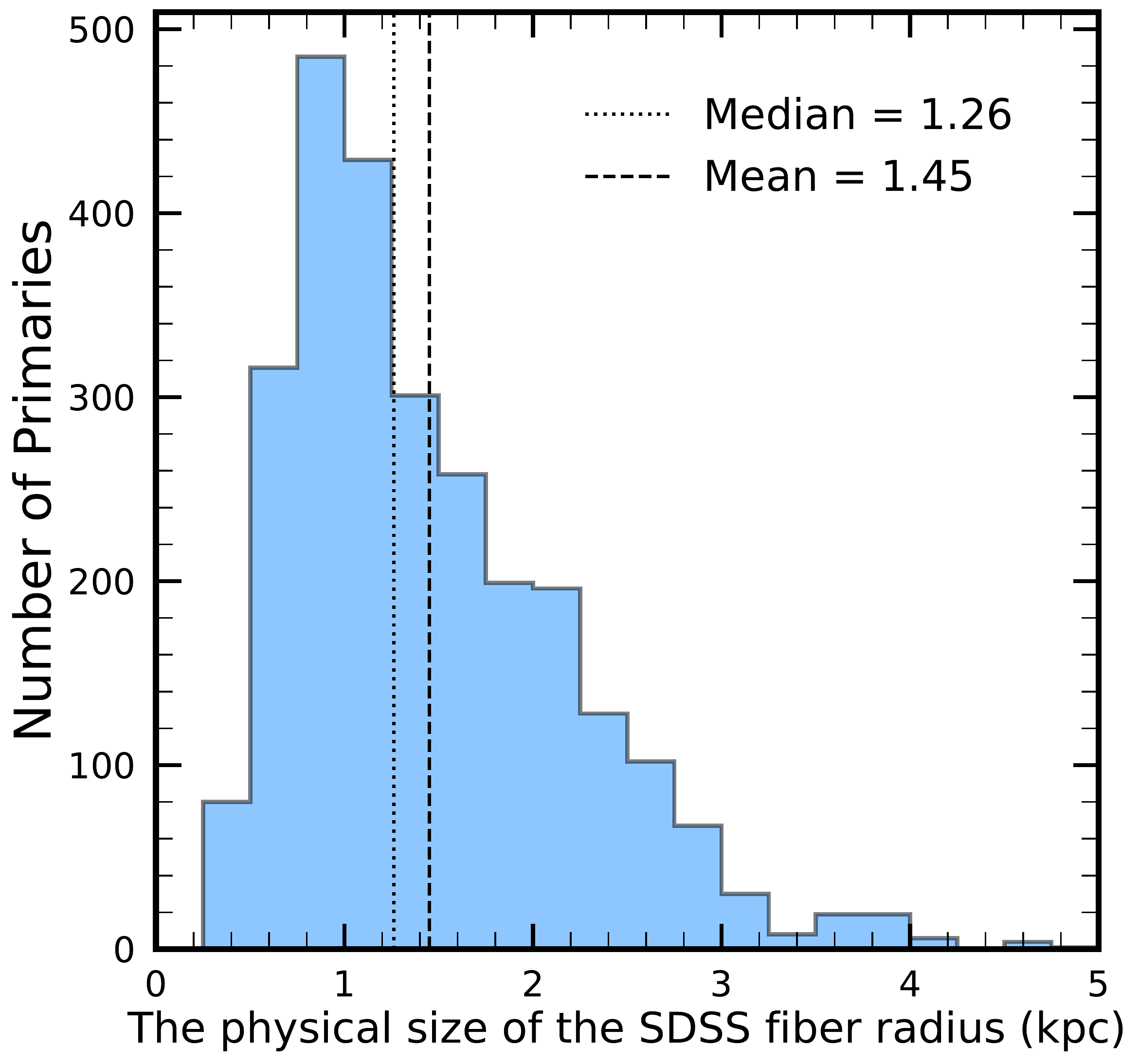}
\caption{A histogram of the linear scale of SDSS fiber radius (1.5 arcsec) used for the central region of each primary galaxy. 
The dashed and dotted lines represent the mean and median value, respectively. }
\label{fig_fiber}
\end{figure}

We obtain the central stellar velocity dispersion of galaxies from the Portsmouth reduction \citep{tho13}. 
These velocity dispersion measurements are in good agreement with those from the SDSS pipeline. 
We use only the values of the galaxies that show 
 a signal-to-noise ratio larger than $\sigma_{0} / \delta \sigma_{0} = 3$.
The central velocity dispersion we use is measured through a 1.5 \arcsec radius SDSS fiber. 
Because our target galaxies are distributed in a wide redshift range $0.01 < z < 0.14$, 
 the physical coverage of the SDSS fiber varies with the redshift of the target galaxies. 
Figure \ref{fig_fiber} shows the distribution of  
 the physical coverage of the SDSS fiber radius on the sample galaxies. 
The physical coverage of the SDSS fiber radius ranges from 0.3 to 5 kpc with a mean value of 1.5 kpc. 

We apply an aperture correction to compute the central velocity dispersion measured with a fixed aperture.
We calculate the aperture-corrected velocity dispersion within 1.5 kpc
 corresponding to the mean SDSS fiber size of the sample.  
The aperture correction we apply is
 $\sigma_{*} / \sigma_{SDSS} = (R_{0} / R_{SDSS})^{-0.054 \pm 0.005}$,
 where $\sigma_{*}$ is the aperture-corrected velocity dispersion, and
 $R_{0}$ is the angular size of 1.5 kpc at given redshifts of galaxies. 
This aperture correction is similar to that in previous studies (e.g., \citealp{cap06, zah16}). 
The typical amount of the aperture correction is only $\sim 2.5\%$. 
Hereafter, we refer to this aperture-corrected velocity dispersion $\sigma_{*}$
 as the central stellar velocity dispersion. 
 
\subsection{Identification of Isolated Galaxies with Satellites}

Field galaxies that host many low-mass satellite galaxies are ideal targets 
 to investigate the relation between the kinematics of satellite galaxies and the dark matter halo. 
However, such systems are rare except for the massive galaxies in the nearby universe \citep{wat10, kar13, kar19}. 
Therefore, most previous studies of satellite kinematics use 
 a method of stacking satellite systems from a large number of isolated galaxies 
 \citep{zar93,zar97,mck02,pra03,bra05,con07,kly09,dut10,mor11,woj13,van16,zah18,lan19}. 
We adopt the same method in this study. 

We construct a complete set of isolated galaxies with faint satellite galaxies from the SDSS spectroscopic sample. 
We use a friends-of-friends (FoF) algorithm \citep{huc82} to find systems hosting a dominant primary galaxy surrounded by faint satellites. 
\citet{soh16} already identified similar systems based on SDSS while they were identifying compact groups of galaxies. 
However, they applied a strict definition of faint satellites  ($\Delta r > 3$ mag) and used a tighter projected linking length of $\Delta D \leq 50 h^{-1}$ kpc. In this study, we extend the identification of such systems in \citet{soh16} based on more generous classification criteria to enlarge the sample size. 

We use a fixed projected linking length of $\Delta D = 100$ kpc and a radial velocity linking length of $|\Delta V | = 1000$ km s$^{-1}$. The radial linking length of $|\Delta V | = 1000$ km s$^{-1}$ is consistent with previous studies that identified galaxy groups \citep{bar96,soh16}. 
We note that more than a half of satellite dwarf galaxies are found at $R < 100$ kpc in the Milky Way galaxy (MWG), which has a virial radius of about 300 kpc \citep{mcc12,kas18,kar19}.
We adopt a projected linking length smaller than the virial radius to reduce any contamination due to interlopers in the samples. 

We then identify isolated galaxies with faint satellite galaxies. We define faint satellite galaxies as the galaxies that are more than 2 mag fainter than their host (i.e., $\Delta r > 2$ mag). We identify the galaxy systems that only contain the faint satellite galaxies. In other words, if an FoF system contains any satellite galaxies with $\Delta r < 2$ mag compared to their host galaxy, we remove this FoF system from our sample. As a result, our sample includes the FoF systems consisting of primary galaxies surrounded by only satellites that are at least 2 mag fainter than their primaries. Because the limiting $r-$band magnitude of the main SDSS spectroscopic sample is 17.77 mag, we identify primary galaxies with $r < 15.77$ mag.
 
We identify 26,430 FoF systems from the SDSS spectroscopic data: 
 22,288 pairs ($N = 2$) and 4142 groups ($N \geq 3$).
Among these systems, there are only 3390 ($\sim 13\%$) systems with 4218 faint satellites that satisfy the satellite magnitude selection ($\Delta r > 2$ mag): 2842 pairs ($N=2$) and 548 groups ($N \geq 3$).
The satellite magnitude selection significantly removes bright pairs and groups that are widely used for studying galaxy properties in galaxy pairs (e.g., \citealp{ell10}) or groups (e.g., \citealp{tem14}). 
We additionally check the contamination of central galaxies in dense cluster environments. 
We compile the three largest catalogs of the BCGs \citep{lau14, klu20, soh20}; 
 there are 566 BCGs within the redshift range $0.00 < z < 0.14$. 
Among the 3390 systems we identify, the primary galaxies of 33 systems are known BCGs. 
To reduce the contamination of the cluster galaxies, 
 we remove these systems (33 primaries hosting 89 satellites) in our sample. 
We then match this list with the catalog of the central stellar velocity dispersion with $\sigma_{0} / \delta \sigma_{0} = 3$. 
The final catalog we use includes 2807 primary galaxies with central stellar velocity dispersion
 and 3417 satellite galaxies.

\begin{figure} 
\centering
\includegraphics[scale=0.32]{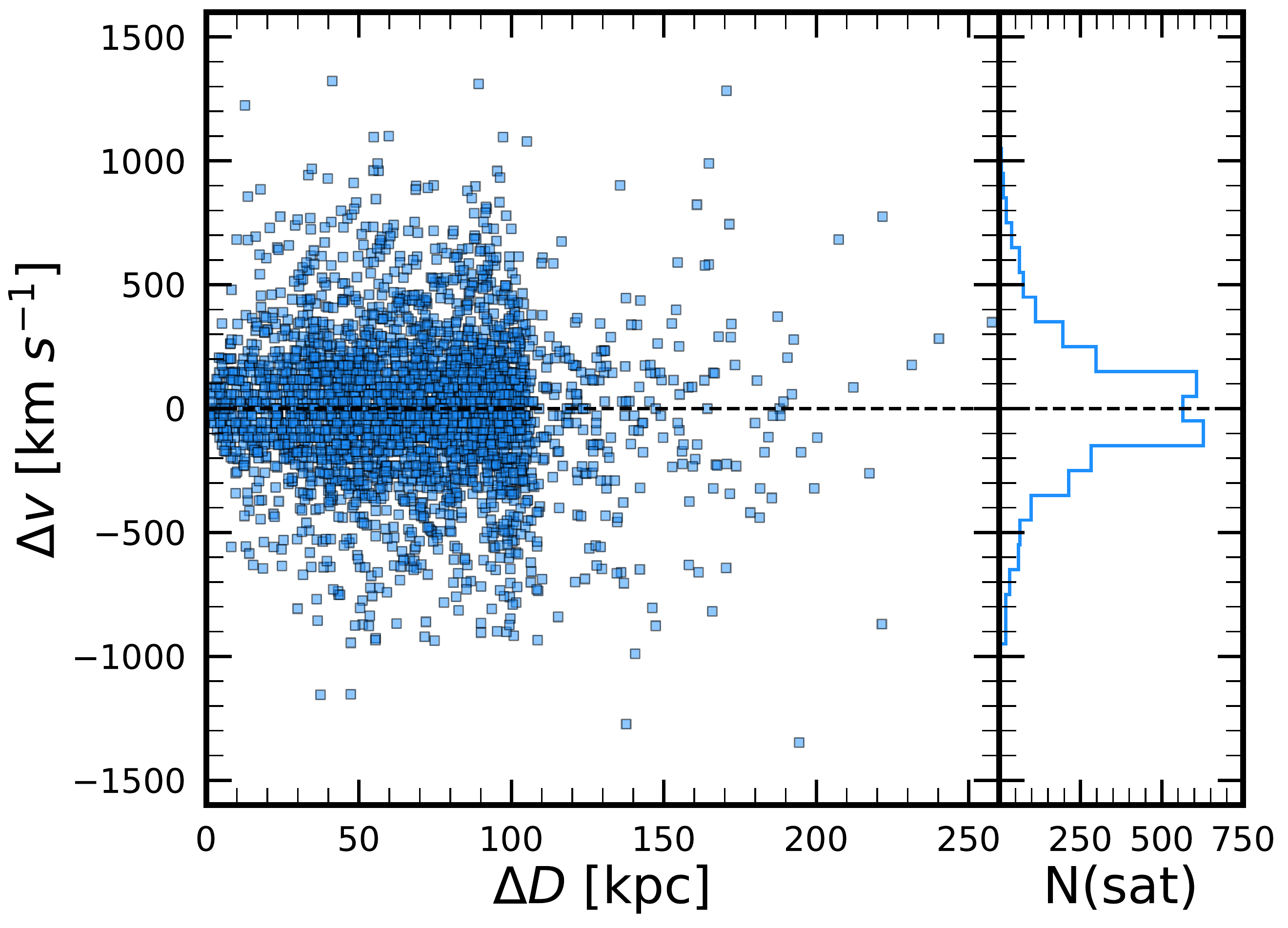}
\caption{
(Left) The relative rest-frame radial velocity differences between primary galaxies and their satellite galaxies 
 as a function of projected galactocentric distance to the satellite galaxies. 
(Right) The distribution of radial velocity differences. }
\label{fig_vdiffD}
\end{figure}

Figure \ref{fig_vdiffD} shows the stacked phase-space diagram (i.e., $R-v$ diagram) of the entire sample. 
We place the primary galaxies at the center and 
 calculate the projected galactocentric distance of satellites ($\Delta D$) and 
 the radial velocity differences ($\Delta V = c \Delta z / (1 + z_{primary})$, 
 where $\Delta z$ is the redshift difference and $z_{primary}$ is the redshift of the primary galaxy). 
A majority of satellites are located at $\Delta D < 100$ kpc, 
 and a small number of them are located out to $\Delta D \approx 260$ kpc.
We plot the histogram of $\Delta V$ for all satellites in the right panel of the figure.  
Most satellites are concentrated around $\Delta V = 0$, which indicates that 
 a majority are gravitationally bound to their primary galaxies.
 
The sharp edge at $\Delta D \sim 100$ kpc is due to the linking length we adopted. 
There are a small number of satellite galaxies at $\Delta D > 100$ kpc. 
These satellites are the members of the systems with $N \geq 3$. 
These satellites are friends of satellites within the linking length from the primary galaxies. 

\begin{figure*} 
\centering
\includegraphics[scale=0.37]{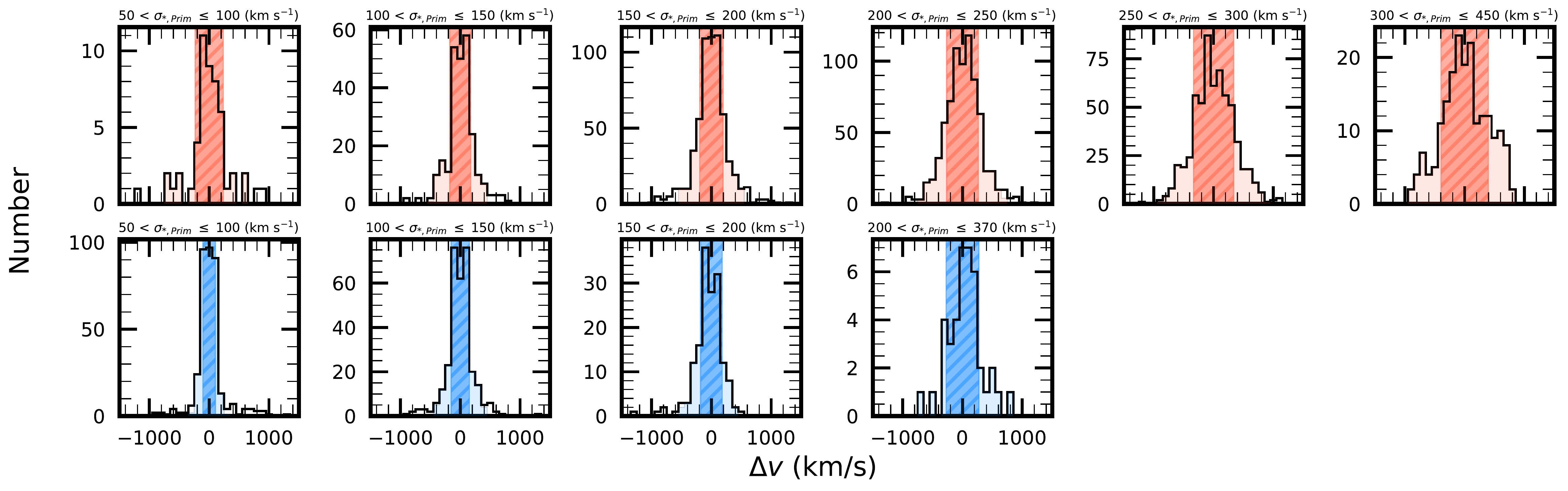}
\caption{The distribution of relative velocity difference ($\Delta V$) of satellite galaxies with respect to their primary galaxies. Top rows show $\Delta V$ distributions for red primary galaxies, and bottom rows show those for blue primary galaxies. Open histograms show the $\Delta V$ of all satellites, light-shaded histograms represent selected satellite galaxies after $2.7\sigma$ clipping, and dark-shaded regions represent the biweight $\sigma_{sat}$ value.}
\label{fig_binhist}
\end{figure*}

The $\Delta V$ distribution shows a trumpet-like shape 
 because the wing components are higher than the normal Gaussian distribution. 
In Figure \ref{fig_binhist}, we display satellite velocity distributions 
 as a function of central stellar velocity dispersion of primary galaxies ($\sigma_{*,prim}$). 
The velocity spread of satellite galaxies increases as $\sigma_{*,prim}$ increases. 
Thus, the trumpet-like shape in Figure \ref{fig_vdiffD} is mainly due to the superposition of narrow velocity distributions of a large number of low-mass systems and broad velocity distributions of a smaller number of high-mass systems, and due to the possibility that the higher velocity dispersion systems (high-mass systems) have proportionally more satellites at larger radii.
 
There are a number of satellite galaxies with $\Delta V > 500$ \kms. 
Those in low-mass systems are mostly $3\sigma$ outliers, while those in high-mass systems are not. 
Those in high-mass systems with $\sigma_{*, prim} > 200$ \kms~ are mostly considered to be satellite members of groups. 

\subsection{Classification of Primary Galaxies}

Previous studies that explore the stellar velocity dispersion of galaxies 
 often select early-type or quiescent galaxies \citep{fab76, zah16, soh17a, soh17b}, 
 using colors or central spectral features $\dn$ of galaxies. 
Here we build two subsamples with early-type and late-type primary galaxies based on color and $\dn$. 
$\dn$ is a spectral index for the strength of the 4000 \AA~ break, 
 which is defined as $F_\nu(4000-4100 {\rm \AA})/F_\nu(3850-3950 {\rm \AA})$ \citep{bal99, fab08, gel14}. 
$\dn$ shows a strong correlation with stellar population age, and  $\dn = 1.6$ corresponds approximately to the stellar population age of 1 Gyr \citep{kau03}. 
In the literature, 
 the galaxies with $\dn \geq 1.6$ are often classified as quiescent galaxies, and 
 those with $\dn < 1.6$ are classified as star-forming galaxies \citep{mig05, woo10}. 
We note that the color of a galaxy represents the mean property of the entire stellar population of the galaxy, 
 while the $\dn$ of a galaxy we measure with SDSS fibers represents only the stellar population in the central region of the galaxy.

\begin{figure} 
\centering
\includegraphics[scale=0.23]{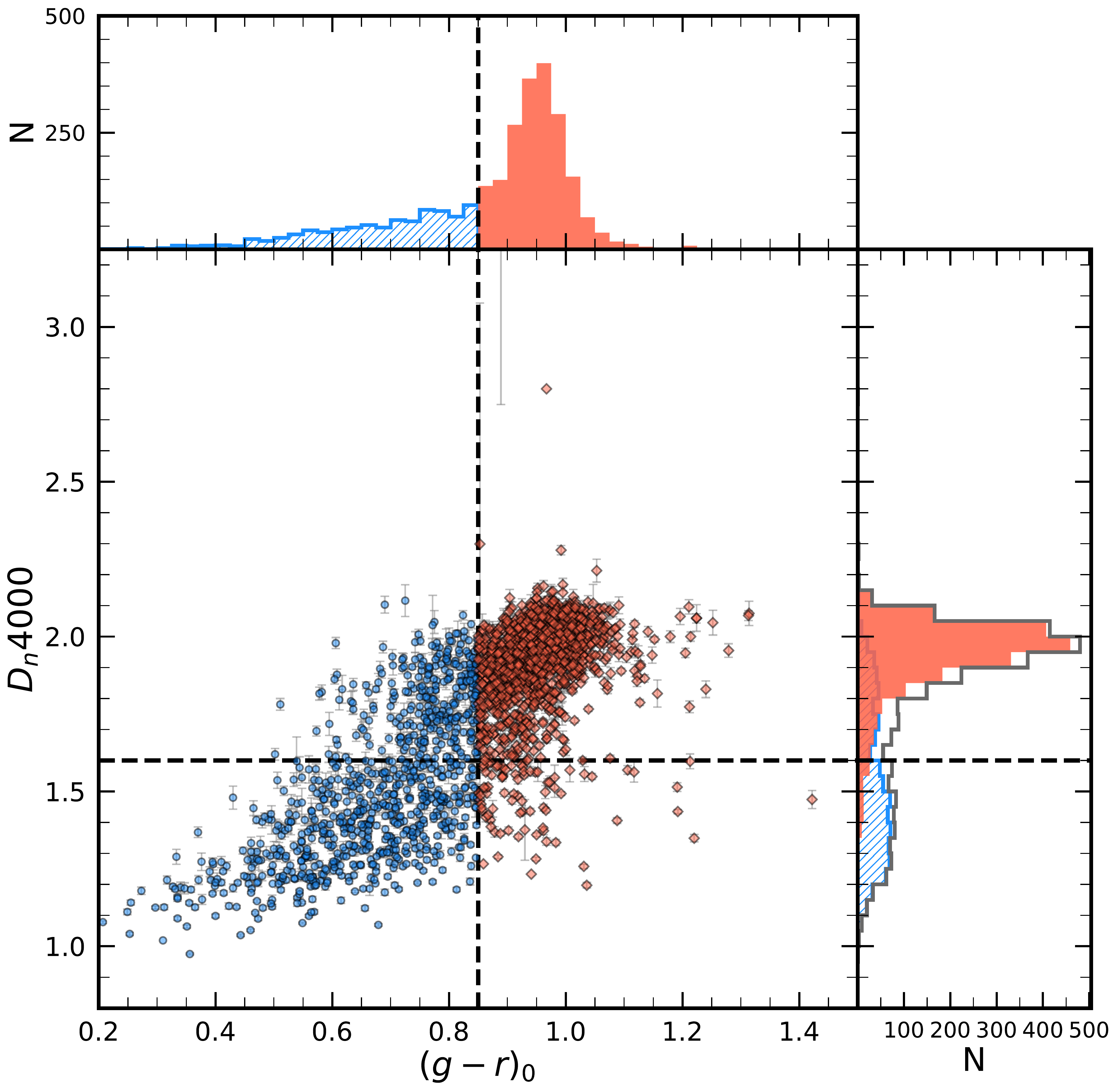}
\caption{$\dn$ vs. $(g-r)_0$ color of primary galaxies and 
 the distribution of each parameter (upper and right panels). 
Blue and red symbols represent blue and red galaxies, respectively.
The dashed lines mark the boundary used for the classification of primary galaxies. }
\label{fig_dn4000color}
\end{figure}

Figure \ref{fig_dn4000color} displays $\dn$ versus $(g-r)_0$ of the primary galaxies. 
$\dn$ shows a strong correlation with $(g-r)_0$ color, but with a significant scatter.
The top panel of Figure \ref{fig_dn4000color} shows 
 the $(g-r)_{0}$ color distribution of the primary galaxies.
The right panel of Figure \ref{fig_dn4000color} displays 
 the histogram of $\dn$ of the primary galaxies.
The $\dn$ distribution of red galaxies shows a dominant single peak at $\dn \approx 2.0$. 
In contrast, the $\dn$ distribution of blue galaxies shows 
 two comparable peaks at $\dn \approx 1.4$ and 1.8, with a minimum at $\dn \approx 1.6$.
This comparison suggests that 
 a quiescent galaxy identification based only on $\dn$ includes a majority of red galaxies, but also includes some blue galaxies. 

\begin{figure} 
\centering
\includegraphics[scale=0.32]{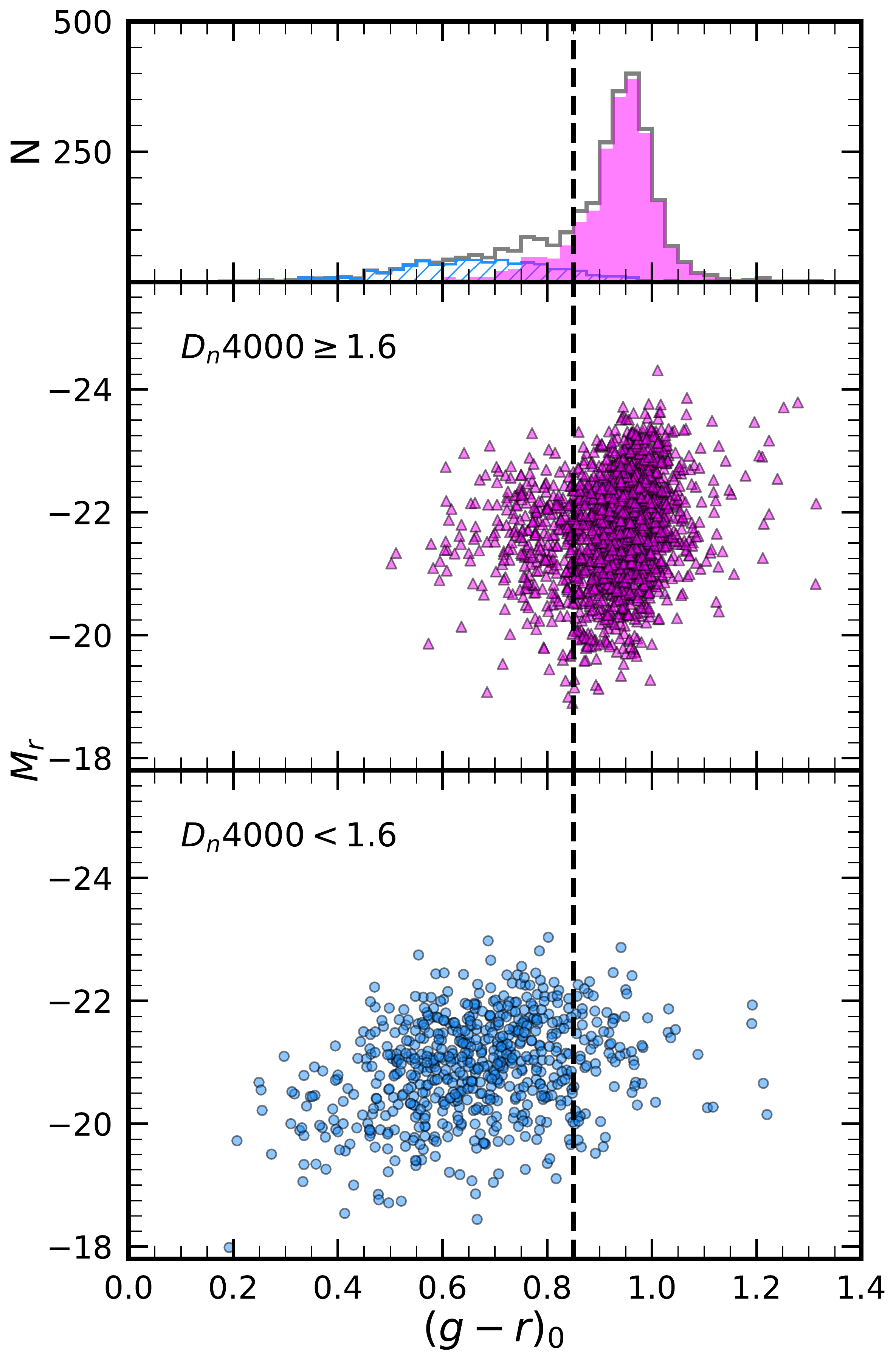}
\caption{Color-magnitude diagrams (middle and bottom panels) and color distributions (top panel) of primary galaxies. 
Red and blue symbols show quiescent and star-forming populations segregated by $D_{n}4000 = 1.6$. 
The black dashed lines mark the boundary of blue and red galaxies. }
\label{fig_cmdDn4000}
\end{figure}

Figure \ref{fig_cmdDn4000} displays 
 a color-magnitude diagram (CMD) of the selected primary galaxies: 
 quiescent galaxies with $\dn \geq 1.6$ and star forming galaxies with $\dn <1.6$.
For comparison, we plot the color histograms of quiescent and star forming galaxies in the top panel. 
The CMD of the quiescent galaxies shows a strong concentration of red galaxies 
 along the red sequence with a peak color at $(g-r)_{0} \approx 0.95$.
Star-forming galaxies are much bluer than the red sequence and show a much broader color range. 
It is also noted that 
 a small number of quiescent galaxies (as well as star forming galaxies) show opposite trends. 
These galaxies must have different star formation histories 
 between the central region and outer region in the same galaxy.

We divide the sample of primary galaxies according to their colors:
 red galaxies with $(g-r)_{0} > 0.85$ and blue galaxies with $(g-r)_{0} \leq 0.85$.
 We divided the sample by color because it is an efficient approximate proxy for morphology (early types vs. late types),
 kinematics (pressure dominated versus rotation dominated) and star-formation history, 
 as adopted in the previous studies (e.g., \citealp{mor11, woj13, lan19}). 
In this classification, 
 $\sim 95\%$ of red galaxies are quiescent galaxies with $\dn \geq 1.6$. 
Based on visual inspection of SDSS images of the primary galaxies, 
 we confirm that the red galaxies are mostly early types and the blue galaxies are mainly late types.
Hereafter, these red and blue primary galaxies are the main subsamples in the following analysis.

We note that we repeated the following analyses based on the quiescent and star-forming galaxies 
 segregated by $\dn = 1.6$. 
The results for quiescent primary galaxies are essentially identical to the results for red primary galaxies. 
Thus, we mention the analyses for quiescent galaxies only when needed in the following.

We include the systems with blue primary galaxies. 
Because blue galaxies are mostly disk galaxies, 
 their central velocity dispersion can be affected by rotation. 
However, the central stellar velocity dispersion of these galaxies measured with an SDSS fiber represents mainly the bulge kinematics.
Thus, it may be much less affected by the effect of rotation; 
 \citet{aqu20} showed that the effect of rotation is minor on the stellar velocity dispersion at $R < 1.5$ kpc in the low-redshift late-type galaxies based on stellar kinematics derived from the MANGA and CALIFA sample of 2458 galaxies (see their Figures 2 and 3). 
We did not take the rotation effect into account 
 in using the measured central velocity dispersion of individual blue galaxies
 (as well as in applying the aperture correction and the inclination correction), 
 because the effect of rotation is considered to be minor and 
 because it is difficult to estimate the fraction of the rotation effect in our sample from the SDSS fiber spectra.

Table \ref{tab_sample} summarizes the number of each subsample. 
Main subsamples for the following analysis are 1938 red primaries with 2461 satellites
 and 869 blue primaries with 956 satellites.
The number of quiescent primaries (2142) with satellites (2681) is slightly larger than that of the red ones. 

\begin{deluxetable}{lcl}
\tabletypesize{\footnotesize}
\tablecaption{Number of Galaxies in Subsamples \label{tab_sample}}
\tablehead{\colhead{Sample} 	& \colhead{$N_{Primary}$} & \colhead{$N_{Satellite}$}}
\startdata
FoF Systems 	                                                          							& 3390  & 4218 \\
All Primaries    ($\sigma_0 / \delta \sigma_0 > 3$)                         	& 2807  & 3417 \\
Red Primaries  ($(g-r)_{0} > 0.85, \sigma / \delta \sigma > 3$)		& 1938  & 2461 \\
Blue Primaries ($(g-r)_{0} \leq 0.85, \sigma / \delta \sigma > 3$)	& 869    & 956
\enddata
\end{deluxetable}

\subsection{Sample Completeness}

The galaxy catalog we use is constructed from the SDSS spectroscopic sample,
 which is slightly incomplete due to fiber collision. 
\citet{laz18} demonstrated that 
 the incompleteness of the SDSS spectroscopic sample is $\sim 7\%$ (see also \citealp{str02}). 
This survey incompleteness may impact the identification of satellite galaxies around isolated galaxies. 

\begin{figure}
\centering
\includegraphics[scale = 0.15]{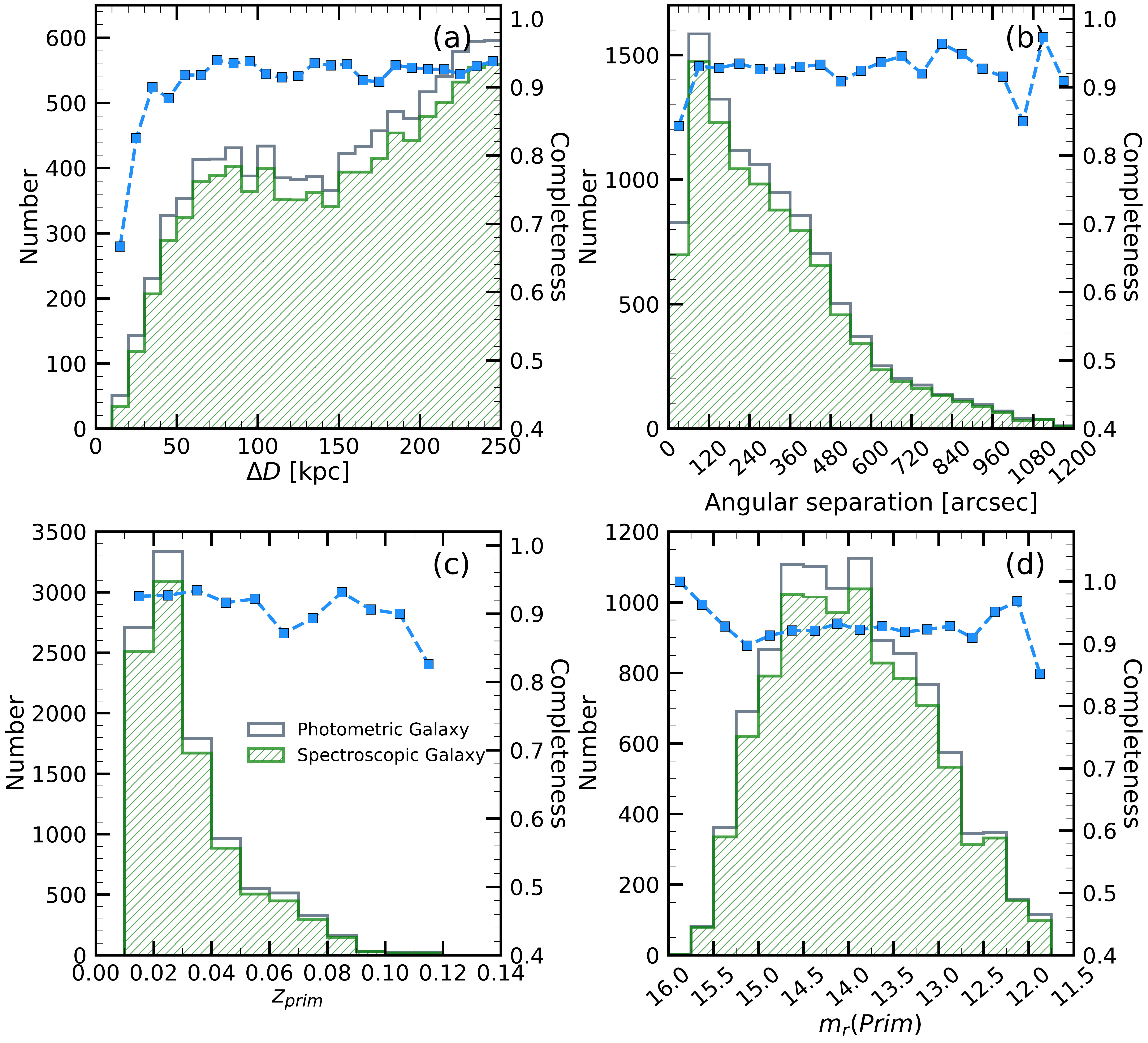}
\caption{The number distribution of faint satellite galaxy candidates (two or more magnitudes fainter than their primary galaxies)  in the photometric (gray open histograms) and spectroscopic (green hatched histograms) samples with respect to 
 (a) the $r$-band magnitudes of primary galaxies, (b) angular separation from primary galaxies, 
 (c) redshifts of primary galaxies, and (d) projected distances of satellites from primary galaxies.
Blue symbols represent the completeness of the spectroscopic samples. 
Completeness scale is labeled on the right axes. } 
\label{fig_completeness}
\end{figure}

We calculate the ratio between the number of spectroscopic galaxies
 and the number of photometric galaxies as a spectroscopic survey completeness. 
Figure \ref{fig_completeness} compares
 the distributions of satellite galaxy candidates in the photometric (open histograms) and spectroscopic (hatched histograms) samples with respect to
 (a) the $r$-band magnitude of primary galaxies,
 (b) angular separation of satellite candidates from their primary galaxies, 
 (c) redshift of primary galaxies, and 
 (d) projected distances ($\Delta D$) of satellites from their primary galaxies.
The blue symbols show the spectroscopic survey completeness. 
Here we compute the spectroscopic survey completeness to $r = 17.77$ mag, the SDSS spectroscopic survey limit. 
The completeness of our sample is larger than 90\% and is almost constant for the entire range 
 of apparent magnitudes, redshifts, and angular separation.
The survey completeness is relatively low ($\sim 70\%$) near primary galaxies ($\Delta D < 20$ kpc), 
 mainly due to the fiber collision. 
However, the number of galaxies with missing spectroscopic information near primary galaxies is also small ($N = 12$).
In conclusion, the spectroscopic survey is essentially complete,
 and it would not significantly affect analysis of the galaxy systems with spectroscopically identified satellites.

\section{Physical Properties of Galaxy Systems} 

\subsection{Satellite Number Distribution}

\begin{figure}
\centering
\includegraphics[scale=0.32]{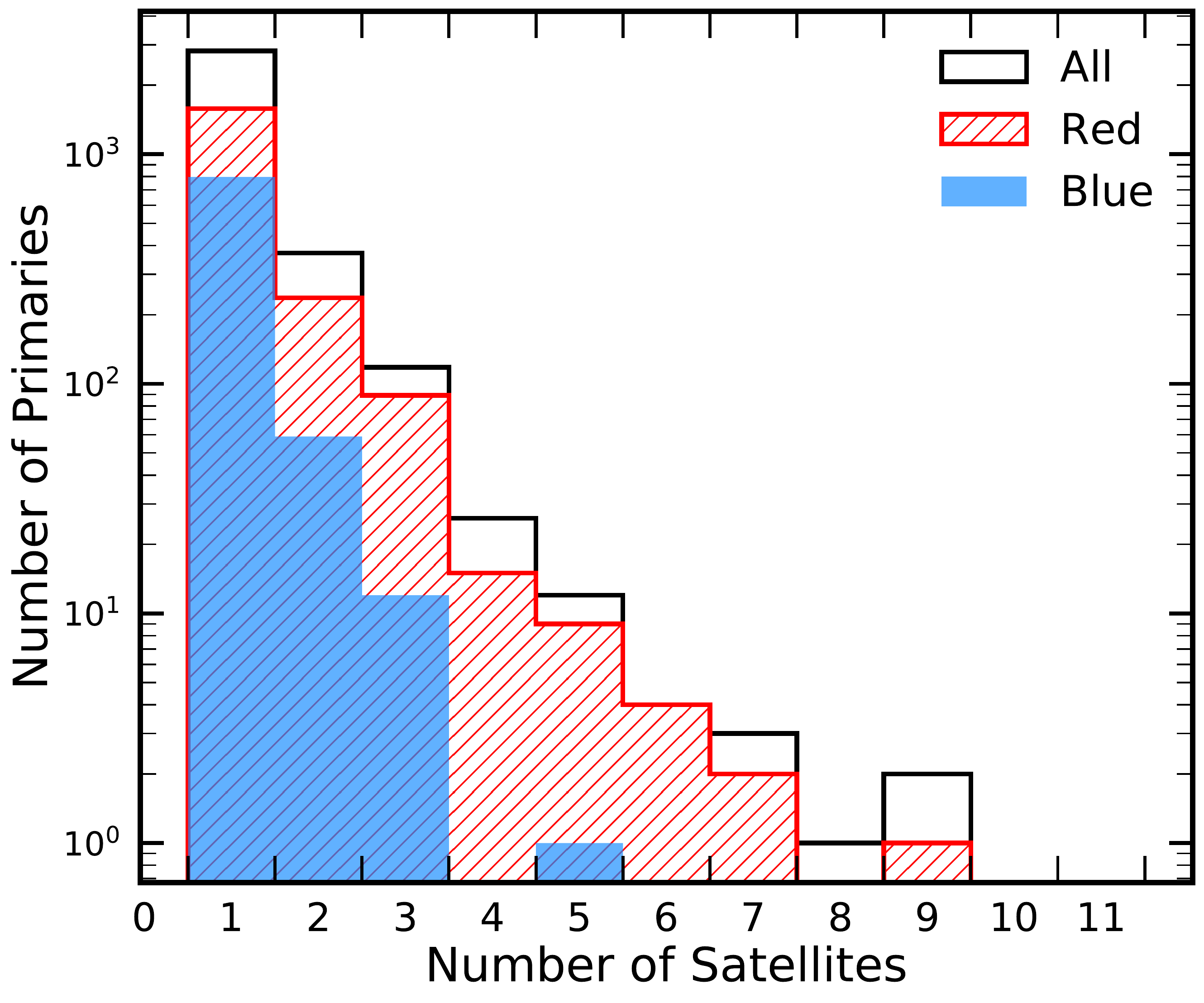}
\caption{Distributions of the number of satellite galaxies for each primary galaxy: 
 all primaries (open), red primaries (hatched), and blue primaries (filled). }
\label{fig_satnum}
\end{figure}

Figure \ref{fig_satnum} displays the distribution of satellite numbers 
 for all (open), red (hatched), and blue (filled) primary galaxies. 
All primary galaxies have fewer than 10 satellite galaxies, and 
 the median number of satellites per primary is 2. 
The blue primary galaxies generally host fewer satellites than the red primary galaxies
 because the former are generally fainter (and less massive) than the latter.
 
\subsection{Estimation of Velocity Dispersion of Galaxy Systems}

The velocity dispersion of relaxed systems is widely used to estimate the dynamical mass of a system based on the virial theorem \citep{wat10}. 
In general, the velocity dispersion of galaxy systems is derived from the line-of-sight radial velocity of satellites except in some cases like the Local group where there are three-dimensional velocities of satellites available. 

Because each primary galaxy in the sample hosts only a few satellites, 
 estimating the velocity dispersion of an individual galaxy system has a significant uncertainty. 
Thus, we compute the system velocity dispersion from the stacked galaxy sample. 
We are interested in the relation between the velocity dispersion of a galaxy system 
 and the central stellar velocity dispersion of the host galaxy in the system. 
Therefore, we stacked samples based on the central stellar velocity dispersion of the host galaxy. 
We divide the red and blue primary galaxies into several velocity dispersion bins with 50 \kms~ intervals.
We also apply $2.7\sigma$ clipping to exclude possible intruders in each sample following \citet{mam10} (see also \citet{fer20} and references therein). 
The line-of-sight velocity dispersion profile measured with the objects within $2.7\sigma$ boundary yields the best fit to that expected from the Navarro-Frenk-White density and velocity anisotropic profile. 
The clipping is iterated until the distribution converges or stops when the iteration number is 5.
In each velocity dispersion bin, there are typically $\sim 480$ (median) galaxy systems with $\sim 590$ (median) satellite galaxies.  
For these samples, we calculate the radial velocity differences of the satellites with respective to the primary galaxies and use them for computing the system velocity dispersion. 

We use the biweight scale estimator \citep{bee90} to calculate the velocity dispersion of galaxy systems.
The biweight method is widely used for computing the velocity dispersion of galaxy systems (e.g., \citealp{zah11, soh20}).
It is known to be more robust than other methods based on the assumption of Gaussian distribution (e.g., \citealp{dan80}), as described in \citet{bee90}. 
We calculate the uncertainties of velocity dispersions using 10,000 times bootstrap resampling. 
Hereafter, we refer to $\sigma_{sys}$ as the system velocity dispersion of the stacked sample.

\section{Results}

Based on the spectroscopically identified sample of isolated galaxies surrounded by faint satellites, 
 we study the relation between the physical properties of the primary galaxies 
 and the properties of their dark matter halo traced by the satellites. 
We first examine the relation between the central stellar velocity dispersion 
 and other physical properties (e.g. luminosity and stellar mass) in Sections \ref{sect_lum} and \ref{sect_mass}. 
We then explore the relations between luminosity, stellar mass, and velocity dispersion of primary galaxies 
 and their system velocity dispersion (Sections 4.3 - 4.5). 

\subsection{Central Stellar Velocity Dispersion and Luminosity of Primary Galaxies}\label{sect_lum}

\begin{figure*}
\centering
\includegraphics[scale=0.58]{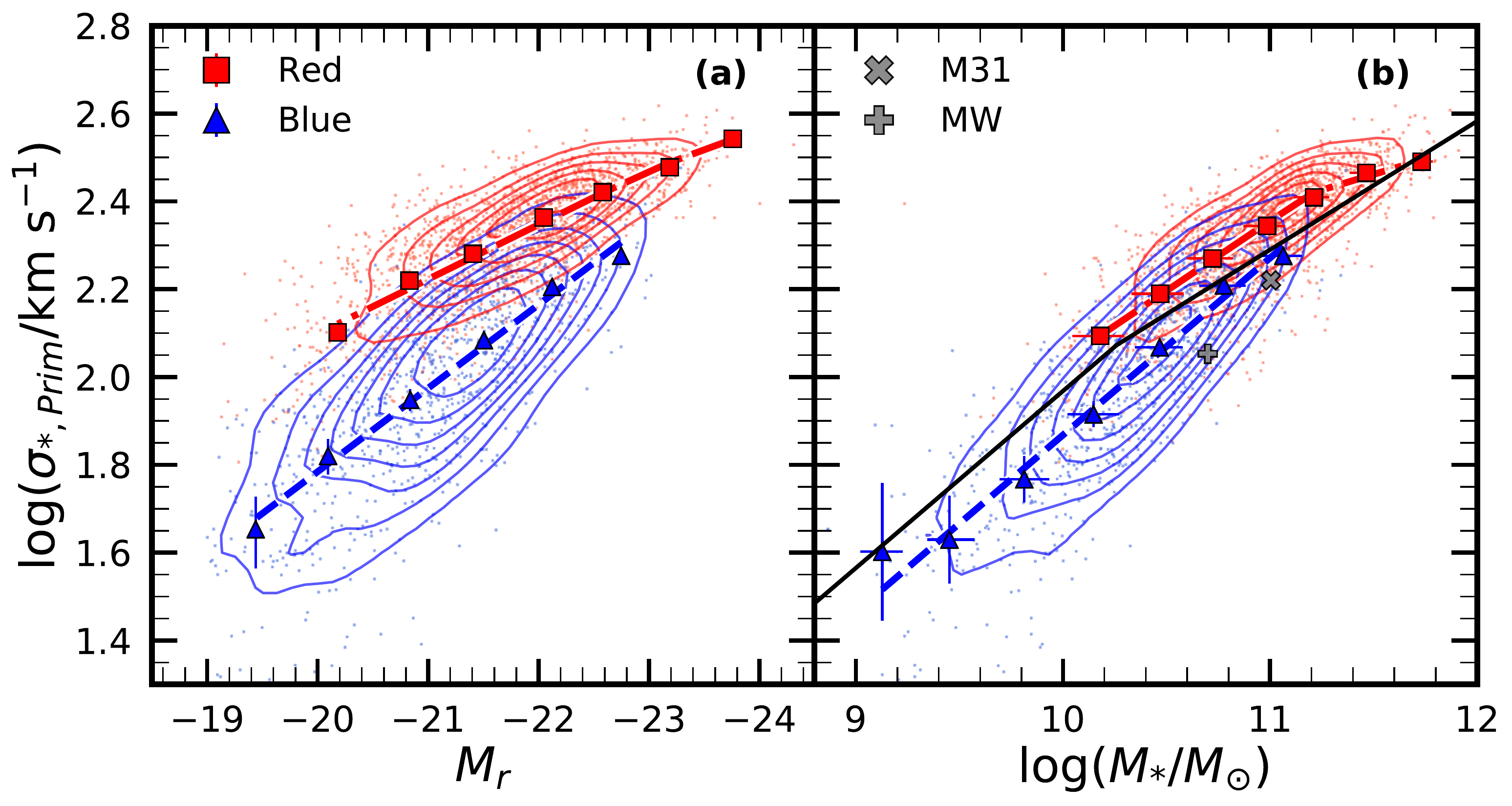}
\caption{Central velocity dispersion as a function of (a) the $r-$band absolute magnitude and (b) stellar mass of primary galaxies. 
Blue and red symbols show the mean values of blue and red galaxies, respectively. 
The blue and red dashed lines are the best-fit relations. 
The black solid line in the right panel shows the relation for SDSS quiescent galaxies in \citet{zah16} (their Table 1).
The gray plus and cross indicate the MWG and M31, respectively. }
\label{fig_vdispmagmass}
\end{figure*}

Figure \ref{fig_vdispmagmass} (a) shows the central stellar velocity dispersion ($\sigma_{*, prim}$)
 as a function of $r$-band absolute magnitude ($M_{r}$) of the blue and red primary galaxies. 
Contours represent the number density of galaxies in the diagram, 
 and the median values of the central velocity dispersion in each bin are marked by the symbols.
In general, there is a strong correlation between $\sigma_{*, prim}$ and $M_{r}$. 
The $\sigma_{*, prim}$ values of the red primary galaxies are, on average, 
 larger than those of the blue primary galaxies for given magnitudes.
The relation for the red primary galaxies shows a break at $M_{r}\approx -23$ mag,  
 and its slope becomes flatter at the brightest end ($M_{r} \leq -23$ mag).

We fit the relations between $\sigma_{*, prim}$ and $M_{r}$ with a linear function. 
Table \ref{tab_cdispmag} lists the fitting results. 
Note that RMS values in these results are as small as $0.01 - 0.02$ dex.
The relation for the red primary galaxies with $M_{r} > -23$ mag is given by  
\begin{equation}
 \log \sigma_{*, prim} = (-0.12 \pm 0.01) M_{r} - (0.29 \pm 0.23), 
\end{equation}
while the relation for the brightest galaxies at $M_{r} \leq -23$ mag is almost flat, 
\begin{equation}
 \log \sigma_{*, prim} = (-0.09 \pm 0.02) M_{r} + (0.40 \pm 0.46). 
\end{equation}
The blue primary galaxies follow a similar relation but with a steeper slope and a $\sim$0.2 dex smaller zero point for given magnitudes:
\begin{equation}
 \log \sigma_{*, prim} = (-0.19 \pm 0.01) M_{r} - (2.02 \pm 0.19). 
\end{equation}
The linear relation of the red primary galaxies suggests $L_{r} \propto \sigma_{*, prim}^{3.3}$,
 which is consistent with the known Faber-Jackson relation. 
This result is also similar to the recent results for the early-type galaxies (with $\log \sigma_{0} = 1.3 -3.2$) in the WINGS survey given by \citet{don20} (see their Figure 1): 
 $\log L_{V} \propto (3.25 \pm 0.07) \log \sigma_{0}$ with an RMS scatter of 0.35 dex.
In contrast, the blue primary galaxies follow $L_{r} \propto \sigma_{*, prim}^{2.1}$ with a smaller power-law index.

\begin{deluxetable*}{clcccl}[h]
\tablewidth{0pc} 
\tablecaption{Fitting results between $M_{r}$ and $\sigma_{*,prim}$ for the sample galaxies \label{tab_cdispmag}}
\tablehead{
\multicolumn{4}{c}{$\log \sigma_{*,prim} = \alpha \times M_{r} + \beta$} \\
\colhead{Sample} 	& \colhead{$\alpha$} & \colhead{$\beta$} & \colhead{Rms [dex]}}
\startdata
All Primaries ($M_{r} > -23$)       & $-0.18 \pm 0.01$	&  $-1.65 \pm 0.23$	& $0.02$  \\
All Primaries ($M_{r} \leq -23$)    & $-0.07 \pm 0.03$	&  $+0.88 \pm 0.69$	& $0.02$  \\
Red Primaries ($M_{r} > -23$)     & $-0.12 \pm 0.01$	&  $-0.29 \pm 0.23$	& $0.01$  \\
Red Primaries ($M_{r} \leq -23$)  & $-0.09 \pm 0.02$	&  $+0.40 \pm 0.46$	& $0.01$  \\
Blue Primaries                               & $-0.19 \pm 0.01$	&  $-2.02 \pm 0.19$	& $0.02$  \\
Satellite                                         & $-0.25 \pm 0.01$	&  $-3.01 \pm 0.18$	& $0.01$
\enddata
\end{deluxetable*}

\subsection{Central Velocity Dispersion and Stellar Mass of Primary Galaxies}\label{sect_mass}

\begin{deluxetable*}{clcccl}[h]
\tablewidth{0pc} 
\tablecaption{Power-law fitting results between $\log(M_{*} / M_{\odot})$ and $\log \sigma_{*,prim}$ for the 
sample galaxies \label{tab_cdispstarmass}}
\tablehead{
\multicolumn{4}{c}{$\log \sigma_{*,prim} = \alpha \times \log (M_{*} / M_{\odot}) + \beta$} \\
\colhead{Sample} 	& \colhead{$\alpha$}    & \colhead{$\beta$} & \colhead{Rms [dex]} }
\startdata
All Primaries ($\log M_{*} \leq 11.2$)    & $0.44 \pm 0.02$	&  $-2.50 \pm 0.22$  & $0.02$  \\
All Primaries ($\log M_{*} > 11.2$)       & $0.14 \pm 0.05$		&  $ +0.86 \pm 0.56$  & $0.01$  \\
Red Primaries ($\log M_{*} \leq 11.2$)  & $0.32 \pm 0.02$	&  $-1.16 \pm 0.22$  & $0.01$  \\
Red Primaries ($\log M_{*} > 11.2$)     & $0.14 \pm 0.02$	&  $ +0.85 \pm 0.22$  & $0.01$  \\
Blue Primaries                              			& $0.40 \pm 0.02$	&  $-2.16 \pm 0.22$  & $0.04$  \\
Satellites                                  				& $0.46 \pm 0.04$	&  $-2.59 \pm 0.38$  & $0.14$
\enddata 
\end{deluxetable*}

In Figure \ref{fig_vdispmagmass} (b), 
 we show $\sigma_{*, prim}$ as a function of stellar mass ($M_{*}$) for the blue and red primary galaxies.
Blue and red symbols show the median $\sigma_{*, prim}$ at each $\log M_{*}$ bin for the blue and red primary galaxies, respectively.  
The $\sigma_{*, prim}$ of the blue and red primary galaxies is tightly correlated with their $M_{*}$, 
 but with a different slope. 
Similar to the relations with luminosity, 
 we note that the slope of red primary galaxies becomes flatter at $\log (M_{*}/M_{\odot}) >11.2$.
This break is consistent with the relation of SDSS early-type galaxies (\citealp{ber11}, see their Figure 1). 
Noting that major dry mergers can change stellar masses of galaxies significantly, 
 but they change the central velocity dispersions (and colors) much less, \citet{ber11} suggest that this break
 (which appeared at $\log (M_{*}/M_{\odot}) = 11.3$ in their case)
 can be evidence of major-dry-merger-driven growth of the massive galaxies. 
 
We fit, with a power law, the median $\sigma_{*, prim}$ of blue and red primary galaxies as a function of $M_{*}$. 
Table \ref{tab_cdispstarmass} summarizes the fitting results. 
The best-fit relation for the red primary galaxies is
\begin{align*}
 &  \log \sigma_{*, prim} = (0.32 \pm 0.02) \log (M_{*}/M_{\odot}) - (1.16 \pm 0.22) \\
 &  \text{~for~} \log (M_{*} / M_{\odot}) \leq 11.2,
\end{align*}
and 
\begin{align*}
 & \log \sigma_{*, prim} = (0.14 \pm 0.02) \log (M_{*}/M_{\odot}) + (0.85 \pm 0.22) \\
 & \text{~for~} \log (M_{*} / M_{\odot}) > 11.2.
\end{align*}
The blue primary galaxies follow a steeper relation with a $\sim$0.1 dex smaller zero point for a given stellar mass,
\begin{equation}
 \log \sigma_{*,prim} = (0.40 \pm 0.02) \log (M_{*}/M_{\odot}) - (2.16 \pm 0.22).
\end{equation}
These relations suggest that the red primary galaxies follow
 $M_{*} \propto \sigma_{*,prim}^{3.1}$, and that the blue primary galaxies follow 
 $M_{*} \propto \sigma_{*,prim}^{2.5}$.
 
We compare the relation for the red primary galaxies with a similar relation derived from SDSS quiescent galaxies from \citet{zah16} 
 (the black solid line in Figure \ref{fig_vdispmagmass} (b)). 
\citet{zah16} use a sample of quiescent galaxies with 
 $\dn > 1.5$, $\log (M_{*} / M_{\odot}) > 9$, and $0.02 < z < 0.20$. 
They also show that the observed relations are fitted with a broken power law; 
 the slope changes at $\log (M_{*} / M_{\odot}) = 10.26$. 
The slope of this relation for the same mass range is similar to the slope of the red primary galaxies we derive. 

We also plot the data for the MWG and M31 in Figure \ref{fig_vdispmagmass} (b).  
We adopt the stellar mass of the MWG ($(5\pm1)\times 10^{10} M_\odot$) and M31 ($1.0\times 10^{11} M_\odot$) from \citet{bla16} and \citet{sic15}, respectively. 
We use the $\sigma_{*, prim}$ of the MWG ($113 \pm 3$ \kms) measured within 
 the half-mass radius of its bulge ($< 1$ kpc) from \citet{bla16}. 
The $\sigma_{*, prim}$ of M31 (173 \kms) is from \citet{whi79} 
(see also a slightly lower value, 166 \kms, in Figure 4 of \citet{sag10}). 
The MWG and M31 are located close to the relation for  the blue primary galaxies.

\begin{figure*} 
\centering
\includegraphics[scale=0.55]{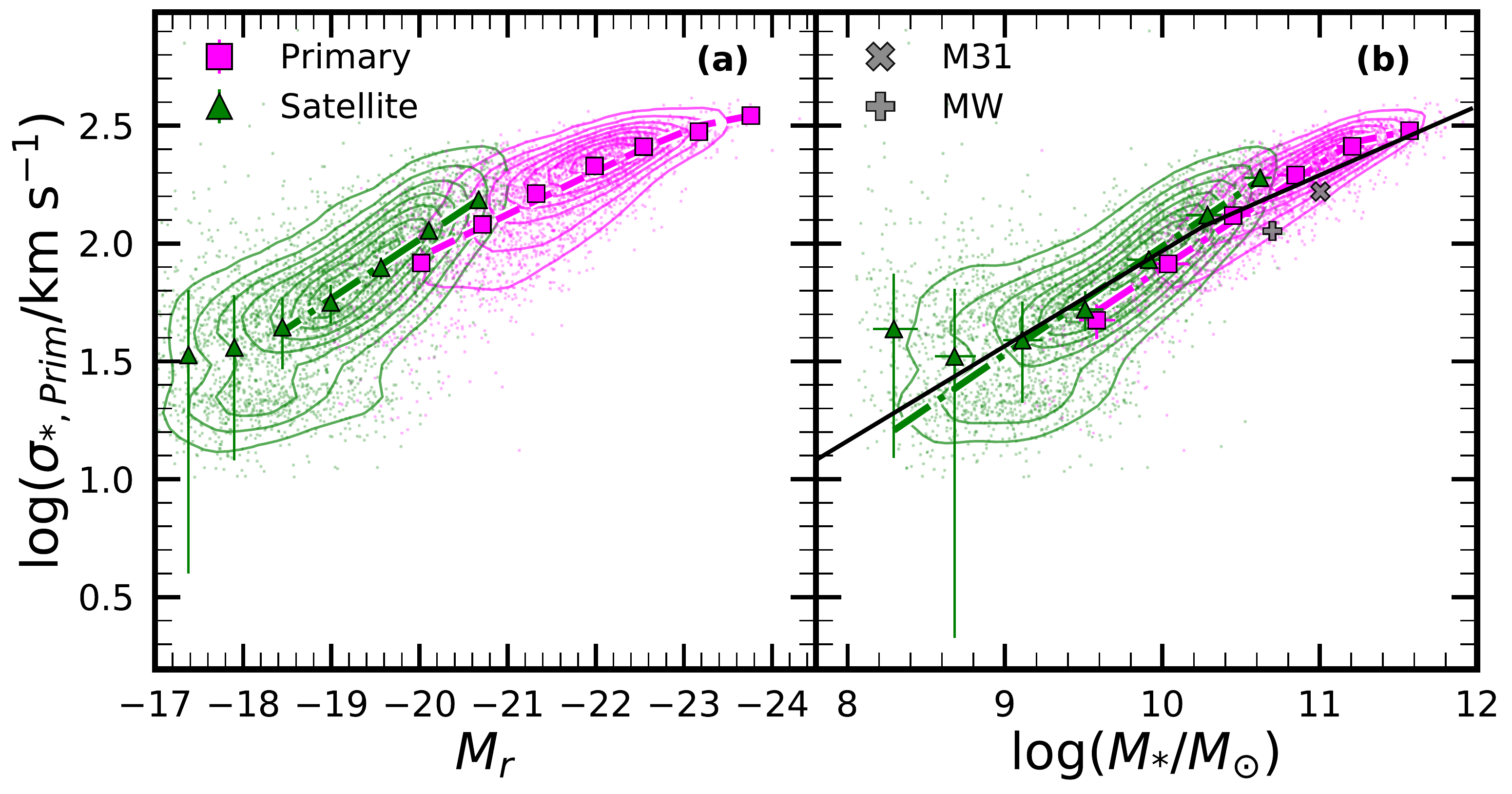}
\caption{Same as Figure \ref{fig_vdispmagmass}, but for primary (magenta) and satellite (green) galaxies. 
The meanings of the symbols are the same as in Figure \ref{fig_vdispmagmass}. } 
\label{fig_vdispmagmassall}
\end{figure*}

Figure \ref{fig_vdispmagmassall} displays $\sigma_{*, prim}$ versus $M_{r}$ and $M_{*}$ 
 for all primary and satellite galaxies.
Fitting results for these data are also listed in Tables \ref{tab_cdispmag} and \ref{tab_cdispstarmass}.
While the primary galaxies show a flatter slope than the satellite galaxies in the $\sigma_{*, prim}- M_{r}$ relation, both galaxy samples show a similar slope in the $\sigma_{*, prim}- M_{*}$ relation.
For comparison we plot the relation for the SDSS quiescent galaxies from \citet{zah16}. 
The relation for the primary and satellite galaxies in this study 
 agrees well with the relation for the quiescent galaxies in \citet{zah16}.
The MWG and M31 are located close to the relation for  the primary galaxies.

\subsection{System Velocity Dispersion versus Luminosity of Primary Galaxies}\label{sect_vdisp}

\begin{figure} 
\centering
\includegraphics[scale=0.35]{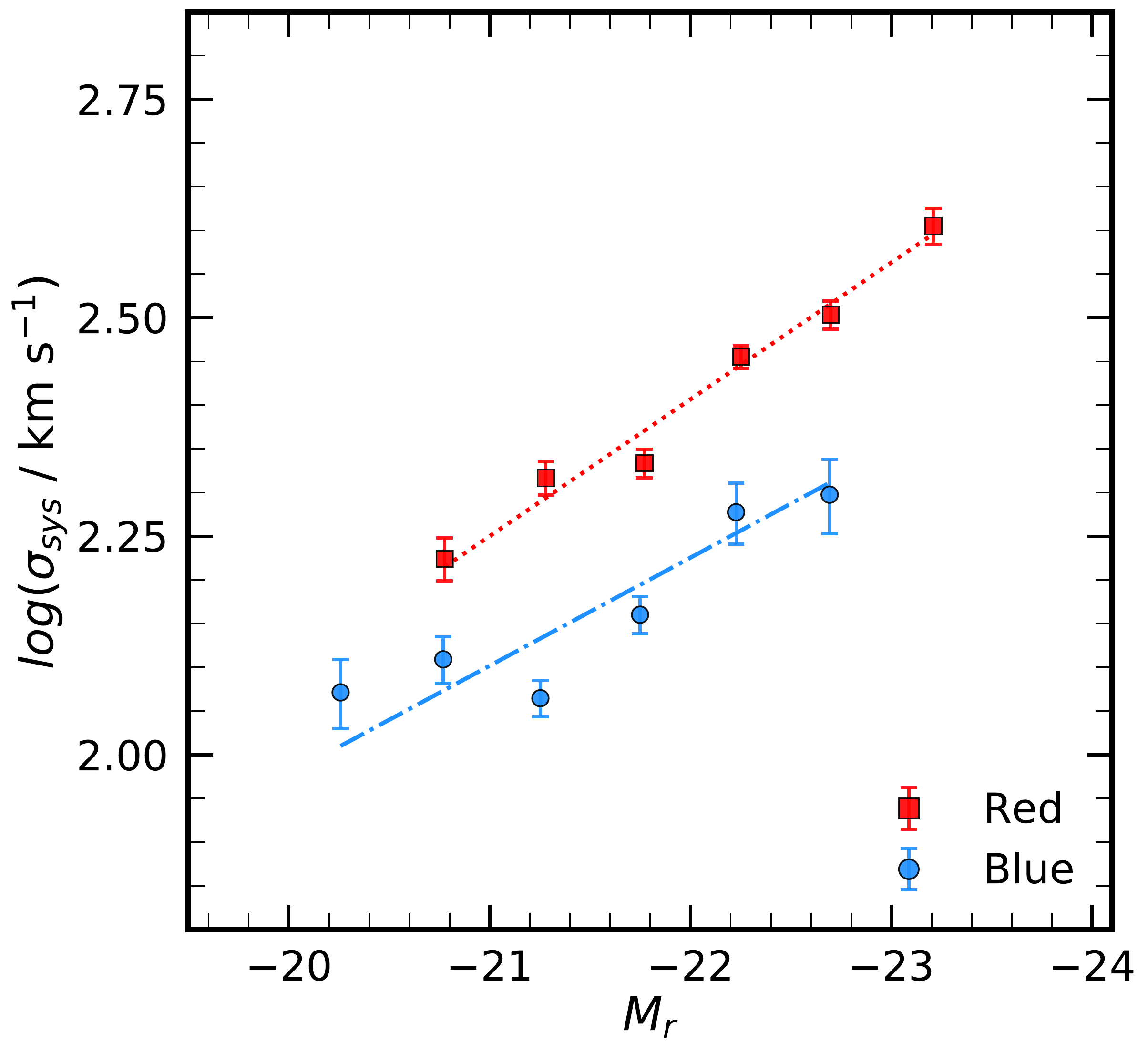}
\caption{Velocity dispersion of galaxy systems as a function of $r-$band absolute magnitude 
 for blue (blue circles) and red (red squares) primary galaxies. 
The blue and red lines show the fitting results for blue and red primary galaxies. }
\label{fig_vdispmag}
\end{figure}

Figure \ref{fig_vdispmag} shows the relation between the system velocity dispersion ($\sigma_{sys}$) 
 and $r-$band absolute magnitude of the primary galaxies. 
Here, we adjusted the bin size along the absolute magnitude so that the number of data in each bin is similar. 
Then, we calculate the system velocity dispersion from the stacked systems. 

For both blue and red primary galaxies, $\sigma_{sys}$ is proportional to $M_{r}$. 
We fit these relations using a linear function. 
Table \ref{tab_vdispmag} lists the best fit results to these relations. 
The best-fit relations are
\begin{align*}
& \log \sigma_{sys} = -(0.16 \pm 0.01) M_{r} -(1.04 \pm 0.27) \\
& \text{~for the red primaries,}
\end{align*}
and
\begin{align*}
& \log \sigma_{sys} = -(0.12 \pm 0.03) M_{r} -(0.50 \pm 0.63) \\
& \text{~for the blue primaries}. 
\end{align*}

Thus, the red and blue primary galaxies show a similar slope in this relation, but the red primary galaxies show, on average, 
 $\sim 0.2$ dex larger $\log \sigma_{sys}$ values than the blue primary galaxies for given magnitudes. 
The best-fit results suggest that 
 the red primary galaxies follow $L_{r} \propto \sigma_{sys}^{2.5}$ and 
 the blue primary galaxies follow $L_{r} \propto \sigma_{sys}^{3.3}$. 
These power-law indices are slightly different from those in the $L_{r} - \sigma_{*,prim}$ relation
 (see Section \ref{sect_lum}). 

\begin{deluxetable}{clcccl}[h]
\tablewidth{0pc} 
\tablecaption{Fitting results between $M_{r}$ and $\log \sigma_{sys}$ for the primary galaxies
\label{tab_vdispmag} }
\tablehead{
\multicolumn{4}{c}{$\log \sigma_{sys} = \alpha \times M_{r} + \beta$} \\
\colhead{Sample} 	& \colhead{$\alpha$} & \colhead{$\beta$} & \colhead{Rms [dex]} }
\startdata
All Primaries		& $-0.15 \pm 0.02$	& $-0.98 \pm 0.50$  & $0.04$ \\
Red Primaries	& $-0.16 \pm 0.01$	& $-1.04 \pm 0.27$  & $0.02$ \\
Blue Primaries	& $-0.12 \pm 0.03$	& $-0.50 \pm 0.63$  & $0.05$ 
\enddata
\end{deluxetable}

\subsection{System Velocity Dispersion and Stellar Mass of Primary Galaxies}\label{sect_mass_relation}

\begin{figure} 
\centering
\includegraphics[scale=0.35]{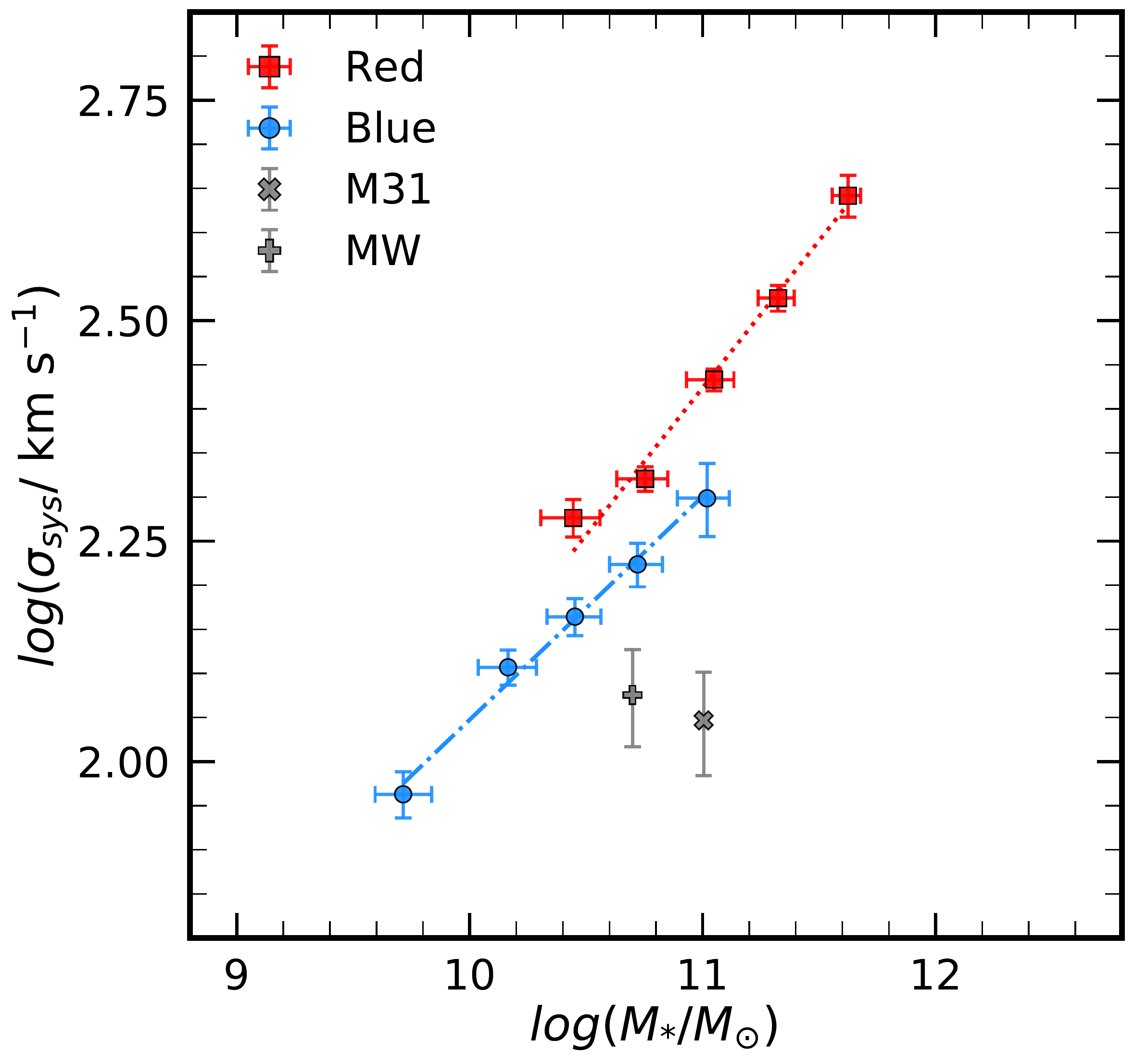}
\caption{Velocity dispersion of galaxy systems as a function of the stellar mass of blue and red primary galaxies.
The lines and symbols are the same as in Figure \ref{fig_vdispmag}.
The gray plus and cross indicate the MWG and M31, respectively. }
\label{fig_vdispmass}
\end{figure}

In Figure \ref{fig_vdispmass} we plot the relation between $\sigma_{sys}$ and the stellar mass of primary galaxies ($M_{*}$).
Similar to the relations with luminosity, 
 $\sigma_{sys}$ is proportional to $M_{*}$ for both red and blue primary galaxies. 
Table \ref{tab_vdispstarmass} lists the best-fit results for these relations we obtain with a power law.
The best-fit relations are
\begin{align*}
& \log \sigma_{sys} = (0.33 \pm 0.03) \log (M_{*} / M_{\odot}) - (1.25 \pm 0.28) \\
& \text{~for the red primaries},
\end{align*}
and 
\begin{align*}
& \log \sigma_{sys} = (0.25 \pm 0.01) \log (M_{*} / M_{\odot}) - (0.49 \pm 0.16) \\ 
& \text{~for the blue primaries}. 
\end{align*}
Thus, the red primary galaxies show, on average, 
 $\sim$0.1 dex larger $\log \sigma_{sys}$ values than the blue primary galaxies for given stellar masses. 
It is noted that the red primary galaxies show a steeper slope than the blue primary galaxies in the $\log \sigma_{sys} - \log M_*$ relation, 
 while they show the opposite trend in the $\log \sigma_{*,prim}- \log M_* $ relation (for $\log M_* < 11.2$).   
These relations indicate that 
$\sigma_{sys} \propto M_{*}^{0.33}$ for the red primary galaxies and 
$\sigma_{sys} \propto M_{*}^{0.25}$ for the blue primary galaxies. 

We plot the data for the MWG and M31 in Figure \ref{fig_vdispmass}.
We estimate the system velocity dispersion of the MWG and M31 using the line-of-sight velocity of their satellite galaxies. 
From the satellite galaxy catalogs in \citet{kas18}, 
 we select 33 and 40 satellite galaxies within 200 kpc from the MWG and M31, respectively. 
Based on their line-of-sight velocities, 
 we compute the system velocity dispersion of the MWG ($\sigma = 119\pm15$ \kms) and M31 ($\sigma_{M31} = 110\pm15$ \kms), respectively.
The errors of the system velocity dispersion were derived by bootstrapping with 1000 times resampling.
The MWG shows a slightly lower $\sigma_{sys}$ value than the mean relation for the blue primary galaxies for given stellar mass, but its offset is only at the 2$\sigma$ level. M31 shows a larger offset at the 3$\sigma$ level.
The line-of-sight velocity distribution of the satellite galaxies of the blue primary systems has a wider range of $-450 < (v_{los} / {\rm km~s^{-1}}) < 450$ than that for the MWG and M31 ($-250 < (v_{los} / {\rm km~s^{-1}}) < 250$). 
The difference in the line-of-sight velocity distribution of satellite galaxies 
appears to be intrinsic because the same difference is shown when we use the blue primary systems identified by a tighter FoF linking length ($\Delta V < 500$ ~\kms). 
The wider radial velocity distribution results in the larger velocity dispersion of the blue primary systems than the MWG and M31. 
To further understand this issue, we need to identify the MWG and M31 system analogs that have a similar magnitude (mass) distribution of the satellite galaxies. 
This is beyond the scope of this paper. 

\begin{deluxetable}{clcccl}[h]
\tablewidth{0pc} 
\tablecaption{Fitting results between $\log(M_{*} / M_{\odot})$ and $\log \sigma_{sys}$ for the primary galaxies
\label{tab_vdispstarmass}}
\tablehead{
\multicolumn{4}{c}{$\log \sigma_{sys} = \alpha \times \log (M_{*} / M_{\odot}) + \beta$} \\
\colhead{Sample} 	& \colhead{$\alpha$} & \colhead{$\beta$} & \colhead{Rms [dex]} }
\startdata
All Primaries		& $0.35 \pm 0.01$	& $-1.42 \pm 0.15$  & $0.02$  \\
Red Primaries	& $0.33 \pm 0.03$	& $-1.25 \pm 0.28$  & $0.02$  \\
Blue Primaries	& $0.25 \pm 0.01$	& $-0.49 \pm 0.16$  & $0.01$
\enddata
\end{deluxetable}

\subsection{System Velocity Dispersion and Primary Central Velocity Dispersion}\label{sect_vdisp_relation}

\begin{figure} 
\centering
\includegraphics[scale=0.32]{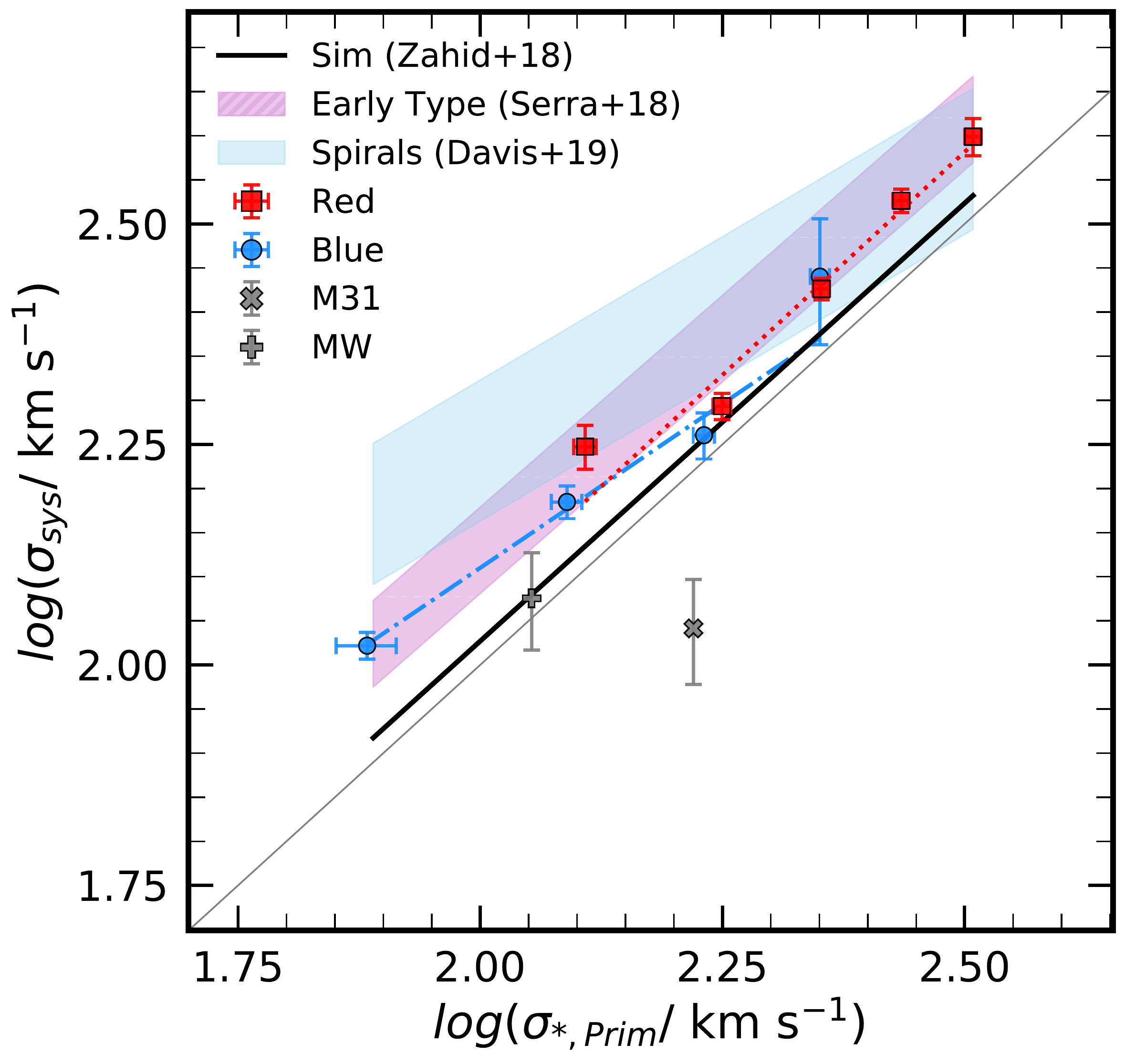}
\caption{
Velocity dispersion of galaxy systems as a function of central stellar velocity dispersion of primary galaxies. 
The symbols are the same as in Figure \ref{fig_vdispmass}. 
The dotted and dotted-dashed lines represent the power-law fits for red and blue primaries.
The black solid line shows the relation for quiescent galaxies from the Illustris-1 simulations \citep{zah18}. 
The cyan band represents the $v_{max}-\sigma_0$ relation for spiral galaxies in \citet{dav19}. 
The magenta band is the $v_{circ}-\sigma_e$ relation for early-type galaxies in \citet{ser16}. 
The gray line is a one-to-one relation.
The gray plus and cross indicate the MWG and M31, respectively. }
\label{fig_vdispcdisp}
\end{figure}

Figure \ref{fig_vdispcdisp} shows the velocity dispersion of galaxy systems ($\sigma_{sys}$)
 versus the central stellar velocity dispersion ($\sigma_{*, prim}$) of the blue and red primary galaxies. 
Remarkably, $\sigma_{*, prim}$ of the red primary galaxies shows a tight correlation with $\sigma_{sys}$.
Table \ref{tab_vdispcdisplog} summarizes the power-law fitting results.
The best-fit result for the red primary galaxies is
\begin{equation}
\log \sigma_{sys} = (1.01 \pm 0.13)~ \log \sigma_{*, prim} + (0.05 \pm 0.31). 
\end{equation}
The slope of this relation is essentially identical to one. 
We also check the same relation for quiescent primary galaxies with $\dn \geq 1.6$; 
 their relation is consistent with that of the red primary galaxies. 
The relation for the blue primary galaxies is slightly flatter ($\alpha = 0.74 \pm 0.09$)
 than for that for the red primaries. 
Nevertheless, the $\sigma_{sys}$ and $\sigma_{*, prim}$ for the blue primary galaxies are 
strongly correlated with each other. 

We compare our results with the relation for the dark matter halos based on hydrodynamic simulations.
Based on the Illistris-1 cosmological simulations, 
 \citet{zah18} derive scaling relations between the dark matter halo mass and the velocity dispersion
 of quiescent galaxies with star formation rates less than $2 \times 10^{-10} M_\odot$ yr$^{-1}$.
Combining their Equations (4) and (7), 
 we obtain a relation between the dark matter halo velocity dispersion ($\sigma_{T,~DM}$) and 
 the line-of-sight stellar velocity dispersion for the half-mass radius of galaxies ($\sigma_{h, *}$):
 $\log \sigma_{T, DM} = 0.99 \log \sigma_{h, *} + 0.06$ with an RMS of 0.17 dex. 
Here we assume that $\sigma_{T, DM}$ corresponds to the system velocity dispersion ($\sigma_{sys}$)
 and $\sigma_{h, *}$ corresponds to the stellar velocity dispersion of primary galaxies ($\sigma_{*, prim}$). 

The black solid line in Figure \ref{fig_vdispcdisp} shows the relation from \citet{zah18}. 
Their relation has a similar slope to that of the red primary galaxies
 although the zero point is slightly lower.
The difference in zero points could have originated from the different definitions of system velocity dispersion in the two studies. 
The relation of the blue primary galaxies is significantly shallower than 
 the relation from the simulations for quiescent galaxies. 
 
We also plot the data for the MWG and M31 in Figure \ref{fig_vdispcdisp}. 
The data for the MWG are consistent with the mean relation for the blue primary galaxies (it is slightly lower but at the $1\sigma$ level). 
However, M31 shows a larger offset at the $3\sigma$ level.
M31 also shows an offset from the simulations \citep{zah18} at the $3\sigma$ level, while the MWG is consistent with the simulations. 

\begin{deluxetable}{clcccl}[h]
\tablewidth{0pc} 
\tablecaption{Power law fitting results between $\sigma_{*, prim}$ and $\sigma_{sys}$ for the primary galaxies} 
\tablehead{
\multicolumn{4}{c}{$\log \sigma_{sys} = \alpha \times \log \sigma_{*, prim} + \beta$} \\
\colhead{Sample} 	& \colhead{$\alpha$} & \colhead{$\beta$} & \colhead{RMS [dex]} }
\startdata
All Primaries				& $0.94 \pm 0.09$	& $0.22 \pm 0.22$  & $0.03$  \\
Red Primaries			& $1.01 \pm 0.13$	& $0.05 \pm 0.31$  & $0.02$  \\
Quiescent Primaries	& $0.99 \pm 0.13$	& $0.10 \pm 0.30$  & $0.03$  \\
Blue Primaries			& $0.74 \pm 0.09$	& $0.62 \pm 0.20$  & $0.01$  \\
\hline
\citet{zah18}			& $0.99 \pm 0.03$    & $0.06 \pm 0.06$ & $0.17$\tablenotemark{a} 
\enddata
\tablenotetext{a}{RMS of \citet{zah18} was derived from quadratic summation of their equations (3) and (8).}
\label{tab_vdispcdisplog} 
\end{deluxetable}

\section{Discussion}

\subsection{Comparison of $\sigma_{sys}-\sigma_{*,prim}$ and $v_{max}-\sigma_0$ Relations}

The rotation velocity of galaxies is another well-known tracer for the dark matter halo of rotating galaxies, 
 which is often applied to disk galaxies \citep{ser16, dav19, kat19}.
For spiral galaxies, the maximum rotational velocity has been used as an accurate tracer of circular velocity at large radii (i.e., $v_{max} \approx v_{circ}$). 
Thus, the maximum rotational velocity is used to estimate the mass of the dark matter halo \citep{dav19, kat19}. 
For example, 
 \citet{dav19} explored a scaling relation between $v_{max}$ and the bulge (central) stellar velocity dispersion ($\sigma_{0}$) based on a sample of 40 spiral galaxies with $\log \sigma_0= 2.0-2.35$.
They showed that there is a correlation between $v_{max}$ and $\sigma_{0}$, but with a large scatter (their equation (6): $\log v_{max} \propto (0.65\pm0.10)\log \sigma_0$ with an RMS of 0.07 dex). 

The cyan band in Figure \ref{fig_vdispcdisp} shows 
 the $v_{max} - \sigma_{0}$ relation for the spiral galaxies from \citet{dav19}. 
The slope of this relation is similar to that of the $\sigma_{sys} - \sigma_{*, prim}$ relation of
 the blue primary galaxies. 
The zero-point difference ($\sim 0.15$ dex) is within in the uncertainties of the relations. 
The consistency in these relations suggest that 
 both $v_{max}$ for the spiral galaxies and $\sigma_{sys}$ for the blue primary galaxies 
 are correlated with the central stellar velocity dispersion of each of the same galaxy. 

Estimating the rotational velocity of early-type galaxies is not straightforward. 
\citet{ser16} measured the rotational velocity ($v_{circ, HI}$) of 16 early-type galaxies 
 hosting a large regular HI disk (or ring). 
They derived a tight relation between $v_{circ, HI}$ and central stellar velocity dispersion 
 ($\log \sigma_{e} = 2.0 - 2.4$):
 $\log v_{max} \propto (0.96\pm0.11) \log \sigma_{e}$ (their equation (1)).
We plot this relation with a magenta band in Figure \ref{fig_vdispcdisp}. 
This relation is in good agreement with the $\sigma_{sys} - \sigma_{*, prim}$ relation for 
 the red primary galaxies we derived. 

\subsection{Relations between Primary Properties and Dark Matter Halo Mass}

Many previous studies investigate the relation between 
 the physical properties of primary (central) galaxies and their dark matter halo mass. 
Following this, we also investigate the relation between the stellar mass and the central velocity dispersion of primary galaxies 
 as a function of the dark matter halo mass inferred from their system velocity dispersions. 
 
We convert $\sigma_{sys}$ into $M_{DM}$ based on the two scaling relations from \citet{rin16} and \citet{abd20}.
\citet{rin16} derive the relation based on the system velocity dispersion of 21 clusters and their mass estimated from Planck Sunyaev-Zel'dovich measurements ($M_{SZ}$, see their equation 5). 
We also used the scaling relation from \citet{abd20} based on GalWcat, 
 which includes $\sim$18,000 galaxy clusters identified from the SDSS spectroscopic sample. 
\citet{abd20} estimate the $\sigma_{sys}$ and $M_{DM}$ of galaxy clusters using a virial mass estimator \citep{bin87, rin13} and derive the best-fit relation $\sigma_{200} = (946 \pm 52) * [h(z) M_{200} / 10^{15} M_{\odot}]^{(0.349 \pm 0.142)}$, where $\sigma_{200}$ is the velocity dispersion of galaxies within $R_{200}$ and $M_{200}$ is the mass within $R_{200}$ in the unit of $[h^{-1} M_{\odot}]$.
The relation explains the observed relation well, particularly at $\sim 10^{14} M_{\odot}$, where the relations derived from cosmological simulations show slight offsets (e.g., \citealp{evr08, mun13, sar13, arm18}). 
We simply assume that $\sigma_{sys}$ and $M_{DM}$ correspond to $\sigma_{200}$ and $M_{200}$ (or $M_{SZ}$), respectively. 

\begin{figure*} 
\centering
\includegraphics[scale=0.55]{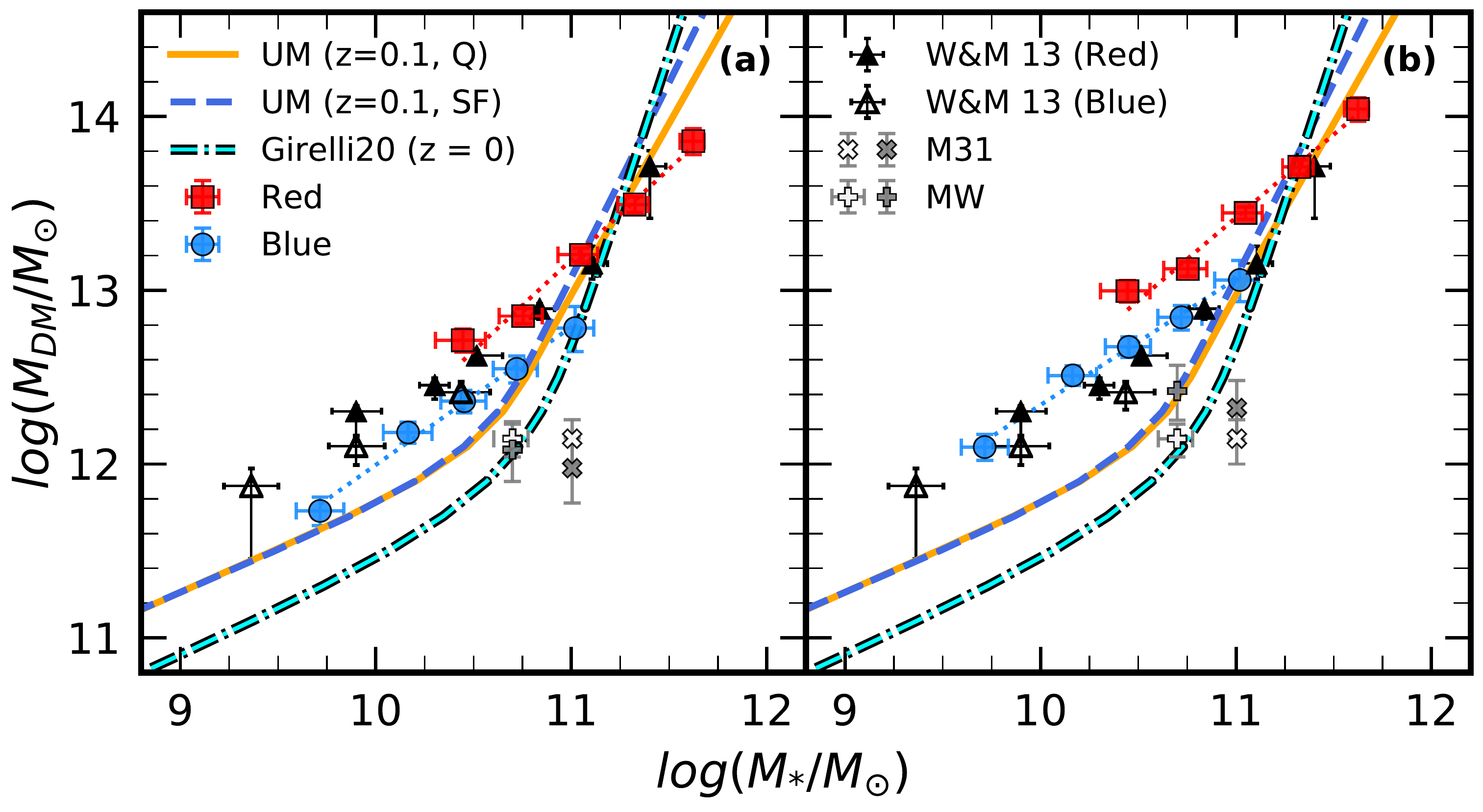}
\caption{Dark matter halo mass as a function of stellar mass of the primary galaxies based on (a) the \citet{rin16} relation and (b) the \citet{abd20} relation. 
Blue and red symbols and lines are for the blue and red primary systems. 
The open plus and cross indicate the MWG and M31, respectively, based on the DM halo mass from the literature.
The filled plus and cross display the MWG and M31, respectively, based on their DM halo mass we translate from their stellar velocity dispersions using the scaling relation. 
Black filled and open triangles show the relations for blue and red SDSS galaxies from \citet{woj13}. 
The relations of \citet{woj13} are shifted to be consistent with the results in this study (see Section 5.2.). 
Yellow solid and blue dashed curves are the relations for quenched and star-formation populations 
 generated with UniverseMachine DR1 \citep{beh19}. 
The dotted-dashed line is the model for $z = 0$ from \citet{gir20}.} 
\label{fig_compwoj13}
\end{figure*}

Figure \ref{fig_compwoj13} illustrates the $M_{*}$ and $M_{DM}$ relations for the isolated galaxy systems. 
We use the scaling relations from \citet{rin16} (the left panel) and \citet{abd20} (the right panel), respectively.
The relations between $M_{*}$ and $M_{DM}$ were derived based on the two scaling relations that show a zero-point offset resulting from the different definitions of $M_{DM}$: $M_{DM} = M_{SZ}$ in \citet{rin16} and $M_{DM} = M_{200, virial}$ in \citet{abd20}.  

There are strong correlations between $M_{*}$ and $M_{DM}$ for both blue and red primary systems, similar to the results from previous observational studies (e.g., \citealp{woj13, erf19}). 
Table \ref{tab_massstarmass} summarizes the fitting results based on the scaling relation from \citet{rin16}
 with the RMS values of $0.04 - 0.05$ dex. 
The best-fit relations between $M_{*}$ and $M_{DM}$ for the red and blue primary galaxies are 
\begin{equation}
 \log (M_{DM} / M_{\odot}) = (1.05 \pm 0.08) \log (M_{*} / M_{\odot}) + (1.66 \pm 0.88), 
\end{equation}
and 
\begin{equation}
\log (M_{DM} / M_{\odot}) = (0.79 \pm 0.05) \log (M_{*} / M_{\odot}) + (4.05 \pm 0.49), 
\end{equation}
respectively. 
We also list the best-fit relations between $M_{*}$ and $M_{DM}$ estimated based on the scaling relation from \citet{abd20} in Table \ref{tab_massstarmass}. 
The slope for the red primary galaxies based on the two scaling relations is essentially identical to one; 
 the slope for the blue primary galaxies is slightly shallower. 
Interestingly, the red primary galaxies are located in a more massive ($\sim$0.4 dex) dark matter halo than the blue primary galaxies at a given $M_{*}$ of the primary galaxies.
 
\begin{deluxetable*}{lcccc}[h]
\tablewidth{0pc} 
\tablecaption{Fitting results between $\log (M_{DM}/M_{\odot}$ and $\log (M_{*}/M_{\odot})$ for the primary galaxies}
\tablehead{
\multicolumn{5}{c}{$\log (M_{DM}/M_{\odot}) = \alpha \times \log (M_{*}/M_{\odot}) + \beta$} \\
\colhead{Sample} 	& \colhead{$\alpha$} & \colhead{$\beta$} & \colhead{Rms [dex]} & \colhead{Ref.\tablenotemark{a}}}
\startdata
All Primaries		& $1.09 \pm 0.04$   & $1.13 \pm 0.46$  & $0.05$ & \\
Red Primaries	& $1.05 \pm 0.08$   & $1.66 \pm 0.88$  & $0.05$ & \citet{rin16} \\
Blue Primaries	& $0.79 \pm 0.05$   & $4.05 \pm 0.49$  & $0.04$ & \\ 
\hline
All Primaries		& $1.00 \pm 0.04$   & $2.41 \pm 0.42$  & $0.04$ &  \\
Red Primaries	& $0.96 \pm 0.07$   & $2.90 \pm 0.80$  & $0.05$ & \citet{abd20} \\
Blue Primaries	& $0.73 \pm 0.04$   & $5.08 \pm 0.45$  & $0.03$ & 
\enddata  
\tablenotetext{a}{References for the satellite velocity dispersion-halo mass relations we used to compute $M_{DM}$. } 
\label{tab_massstarmass}
\end{deluxetable*}

Figure \ref{fig_compwoj13} compares the relations we derive with those from \citet{woj13},
 also obtained from the isolated galaxy systems in the SDSS spectroscopic sample. 
\citet{woj13} identified the isolated galaxy systems from SDSS DR7 based on more generous criteria than our selections: 
 $\Delta D < 1$ Mpc, $|\Delta V| < 1500$ km s$^{-1}$, and $\Delta M_{r} > 1.505$. 
Their sample includes 3800 red and 1600 blue primary galaxies surrounded by 8800 and 2600 satellite galaxies, respectively. 
The red and blue primary galaxies are separated based on the color-magnitude relation rather than a simple color selection. 
They obtained the stellar mass of galaxies from the SDSS MPA/JHU catalog \citep{sal07}; 
 these measurements are systematically larger ($\sim$0.2  dex) than the mass estimates we use. 
Thus, we shifted the relations from \citet{woj13} by $-0.2$ dex in the $M_{*}$ direction
 for a fair comparison. 
They compute the $M_{DM}$ values using their projected phase-space (PPS) model, 
 which yields an $M_{DM}$ distribution consistent with that from $\Lambda$CDM simulations.

The $M_{*}$ and $M_{DM}$ relations we derive using the \citet{rin16} relation are consistent with the results from \citet{woj13}, while those we derive using the \citet{abd20} relation show slight offsets to the higher $M_{DM}$. 
We note that the $M_{DM}$ used in \citet{woj13} is slightly lower than the $M_{DM}$ used in other previous studies
 (see discussion in \citealp{woj13, van16}). 
For example, \citet{van16} discussed that 
 the $M_{DM}$ from \citet{woj13} is $30\% - 40\%$ lower than the mass estimates based on weak lensing analysis. 

For comparison, we show the MWG and M31 in Figure \ref{fig_compwoj13}.
We derive $M_{DM, MW}$ from $\sigma_{sys}$ values for the MWG and M31 using the scaling relations in \citet{rin16} and \citet{abd20} and plot them with filled symbols.
We also obtain $M_{DM, MW}$ ($(1.4\pm0.3)\times 10^{12} M_{\odot}$) and $M_{DM, M31}$($(1.4\pm0.4) \times 10^{12} M_{\odot}$) estimated from \citet{wat10}, plot them with open symbols. 
The values we derive using the \citet{rin16} relation are more similar to those based on \citet{wat10}.
The MWG and M31 show offsets to the lower value at the $2\sigma$ and $3\sigma$ levels from the relation for the blue primary systems derived using the \citet{rin16} relation, and they show slightly larger offsets from the results based on the \citet{abd20} relation.

We also compare the observed relations with similar relations 
 based on simulations from UniverseMachine DR1 \citep{beh19} and 
 on models from \citet{gir20}.
The solid and dashed lines in Figure \ref{fig_compwoj13} display the 
 $M_{*} - M_{DM}$ relation for quenched and star-forming populations for $z=0.1$, respectively, 
 in the UniverseMachine. 
Because UniverseMachine does not provide models for blue and red galaxies separated by colors, 
 we assume that the models for quenched and star-forming populations 
 correspond to red and blue galaxies in this study.
 
There are interesting differences between the relations from observations and the UniverseMachine.  
The UniverseMachine models show a break at $\log (M/M_\odot) \approx 10.5$, 
 which is absent from our results (also from the results of \citealp{woj13}). 
The UniverseMachine models at $\log (M_{DM}/M_{\odot}) > 12$ are much steeper than the relations 
 for the red primary galaxies in this study. 
The relation for $z=0$ in \citet{gir20} is even steeper than the UniverseMachine models at the high-mass end.
In addition, the difference between the quenched and star-forming models in UniverseMachine is 
 much smaller than that between the red and blue primary galaxies in this study.
Further studies with independent measurements of $M_{DM}$ of the galaxy systems (e.g., based on lensing techniques) are needed to understand these differences with the models. 

\begin{figure*}
\centering
\includegraphics[scale=0.55]{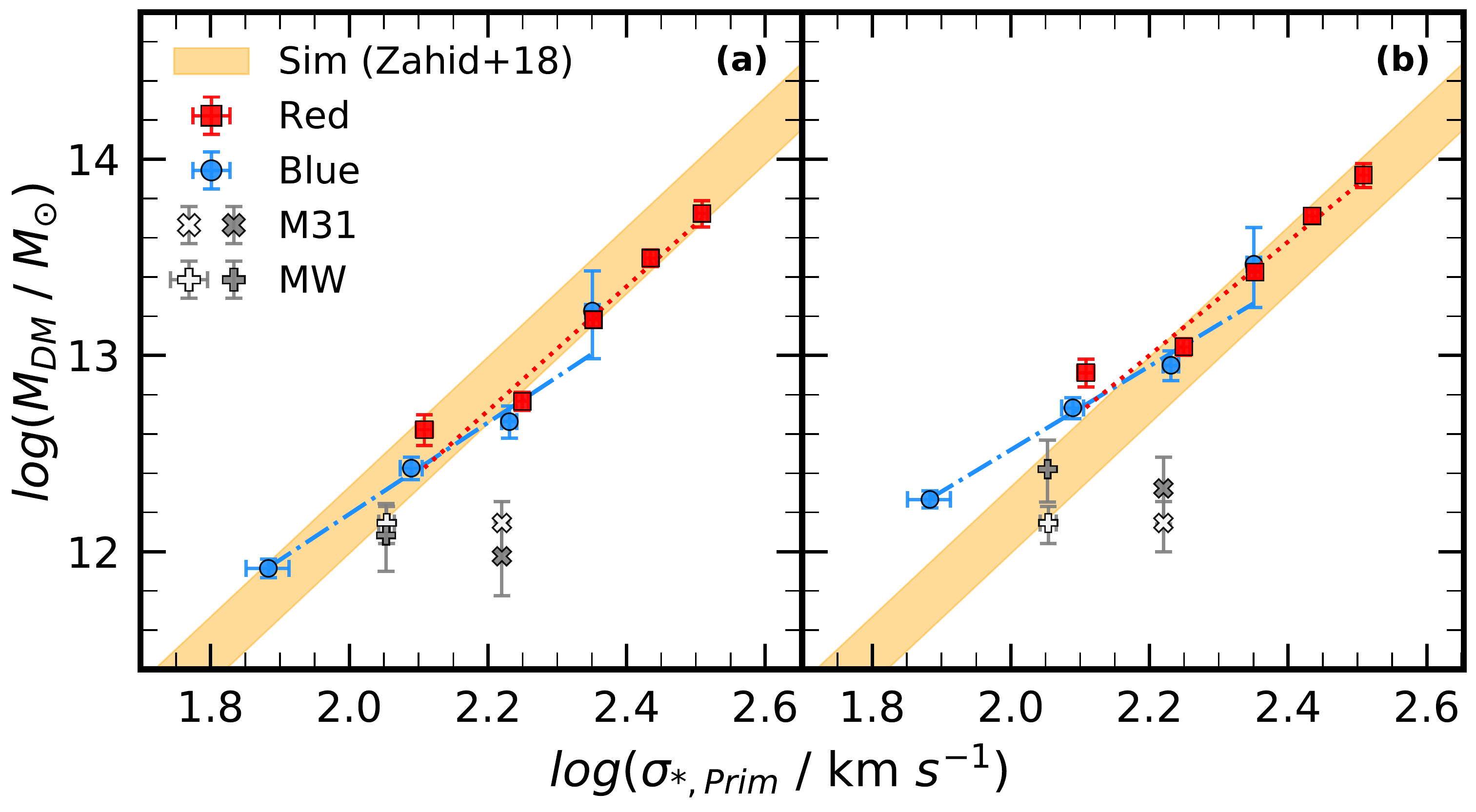}
\caption{Dark matter halo mass (derived from satellite velocity dispersion using the relation in (a) \citet{rin16} and (b) \citet{abd20}) vs. the central velocity dispersion of blue and red primary galaxies. 
The yellow band represents the relation ($M_{200}$ vs. $\sigma_{h,*}(=\sigma_{*,prim}$)) based on Illustris-1 simulations given by \citet{zah18}. 
The meanings of the gray symbols are the same as in Figure \ref{fig_compwoj13}.} 
\label{fig_compzahid18}
\end{figure*}
 
Figure \ref{fig_compzahid18} shows the $M_{DM}$ we derived as a function of $\sigma_{*, prim}$. 
As expected from the linear relation between $\sigma_{*, prim}$ and $\sigma_{sys}$ derived in this study,
 $M_{DM}$ is proportional to $\sigma_{*, prim}$. 
We fit the relation we derive using the scaling relation from \citet{rin16} and \citet{abd20} with a power law,
 listing the results in Table \ref{tab_csigmahalomass}.
The rms values are 0.04--0.10. 
The best-fit relations between $\sigma_{*,prim}$ and $M_{DM}$ estimated using the scaling relation from \citet{rin16} are 
\begin{equation}
 \log (M_{DM} / M_{\odot}) = (3.17 \pm 0.41) \log \sigma_{*,prim} + (5.74 \pm 0.97), 
\end{equation}
and 
\begin{equation}
 \log (M_{DM} / M_{\odot}) = (2.33 \pm 0.29) \log \sigma_{*,prim} + (7.53 \pm 0.61), 
\end{equation}
for the red and blue primary galaxies, respectively. 
These relations are useful in estimating the dark matter halo mass of the galaxies 
 for which the values of the central velocity dispersion are available.

The relation for the red primary galaxies based on the \citet{rin16} relation 
agrees very well with the expected relation 
 based on the Illustris-I simulations \citep{zah18}, as shown by the yellow band. 
The observed relation based on the \citet{abd20} scaling relation shows an offset toward high $M_{DM}$ compared to the relations from the numerical simulations.
The data of the MWG are consistent with the relation for the blue primary galaxies, and M31 shows an offset at the 3$\sigma$ level to the lower $M_{DM}$.

\begin{deluxetable*}{lcccc}[h]
\tablewidth{0pc} 
\tablecaption{Fitting results between $\sigma_{*, prim}$ and $M_{DM}^a$ for the primary samples}
\tablehead{
\multicolumn{5}{c}{$\log M_{DM} = \alpha \times \log \sigma_{*, prim} + \beta$} \\
\colhead{Sample} 	& \colhead{$\alpha$} & \colhead{$\beta$} & \colhead{Rms [dex]} & \colhead{Ref.\tablenotemark{a}}}
\startdata
All Primaries         & $2.94 \pm 0.29$   & $6.27 \pm 0.68$ & $0.10$ & \\
Red Primaries       & $3.17 \pm 0.41$   & $5.74 \pm 0.97$ & $0.08$ & \citet{rin16} \\
Blue Primaries      & $2.33 \pm 0.29$   & $7.53 \pm 0.61$ & $0.04$ & \\
\hline
All Primaries         & $2.69 \pm 0.27$   & $7.11 \pm 0.62$ & $0.09$ & \\
Red Primaries       & $2.90 \pm 0.38$   & $6.63 \pm 0.89$ & $0.07$ & \citet{abd20} \\
Blue Primaries      & $2.13 \pm 0.27$   & $8.26 \pm 0.56$ & $0.09$ & 
\enddata
\tablenotetext{a~}{References for the $M_{200} - \sigma_{sys}$ relations we used to compute $M_{DM}$. }
\label{tab_csigmahalomass}
\end{deluxetable*}

\subsection{Comparison with Massive Clusters} 

We demonstrated that there is a clear relation between $\sigma_{*, prim}$ and $\sigma_{sys}$
 for both blue and red primary systems. 
\citet{soh20} investigate a similar relation between the stellar velocity dispersion of 
 the BCGs and the cluster velocity dispersion 
 derived from a spectroscopic sample of cluster members. 
They suggest that this relation is an important test for central galaxy formation and structure formation models. 

\begin{figure*} 
\centering
\includegraphics[scale=0.65]{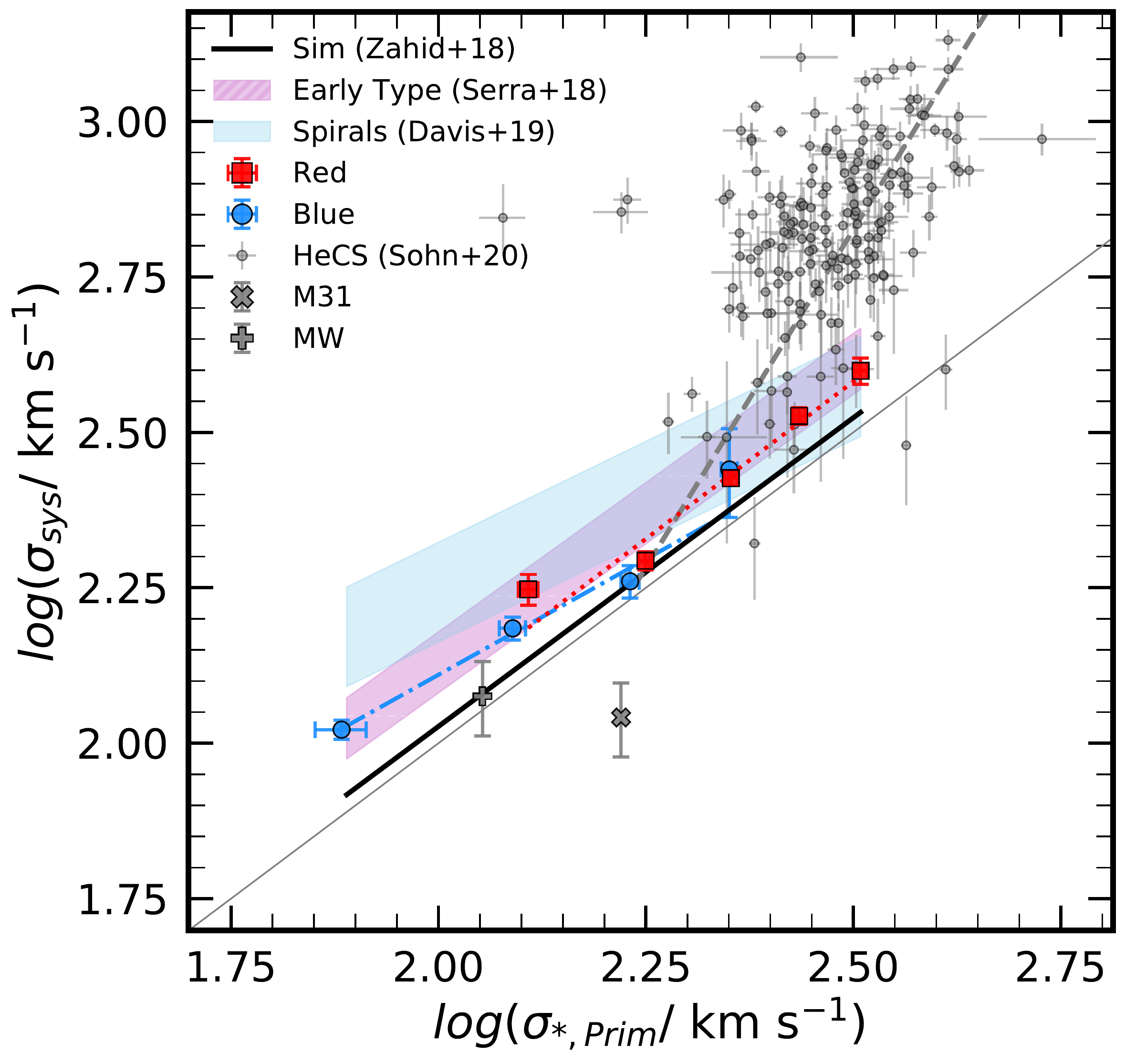}
\caption{Same as Figure 11, but in comparison with the HeCS-omnibus galaxy cluster sample from \citet{soh20}. 
The symbols are the same as in Figure \ref{fig_vdispcdisp}. 
Gray points are the individual galaxy clusters, and the  
gray dashed line shows the best-fit relation for these galaxy clusters given by \citet{soh20}.
The gray plus and cross indicate the MWG and M31, respectively.}
\label{fig_vdispcdisp_bcg}
\end{figure*}

Figure \ref{fig_vdispcdisp_bcg} compares the $\sigma_{sys} - \sigma_{*, prim}$ relations 
 for the isolated galaxy systems in this study and for the cluster sample (gray circles) from \citet{soh20}. 
The comparison cluster sample includes 225 HeCS-omnibus clusters. 
The velocity dispersions of the clusters ($\sigma_{cl}$) are derived from spectroscopically identified cluster members ($\sim$180 for each cluster) using the biweight scale, as used in this study.
A majority of the HeCS-omnibus clusters have $400 <\sigma_{cl} <1000$ \kms, 
 much larger than the $\sigma_{sys}$ for the isolated galaxy systems in this study. 
The gray dashed line is the best-fit result from \citet{soh20}. 

Interestingly, the relation we derived for the isolated galaxy systems in this study differs 
 clearly from the relation for the galaxy clusters. 
Although the BCGs in some clusters with small $\log \sigma_{cl} \sim 2.5$ have similar stellar velocity dispersions to those of the isolated systems, a majority of the galaxy clusters show a much steeper $\sigma_{cl} - \sigma_{*, prim}$ relation. 
The relation for galaxy clusters also has a different slope 
 compared to the relation from the numerical simulation in \citet{zah18}, as \citet{soh20} pointed out. 
 
We note that the definitions of $\sigma_{sys}$ in the two studies are different. 
For isolated galaxy systems, 
 we use satellite galaxies identified by the FoF algorithm with the limited linking length, 
 which may include some intruders. 
In the case of galaxy clusters, the cluster members are identified based on the caustic technique \citep{dia97}.
Furthermore, we compute $\sigma_{sys}$ from the stacked sample including many isolated systems, 
 while \citet{soh20} measured $\sigma_{cl}$ for individual galaxy clusters. 
\citet{zah18} estimate the stellar velocity dispersions and 
 the dark matter halo velocity dispersions from the simulations in a similar manner to this study. 
This difference in definition would introduce some differences in the relations. 

The different relations of the galaxy clusters and the isolated galaxy systems indicate that 
 $\sigma_{sys}$s for these two systems may trace different halos. 
In other words, the $\sigma_{sys}$ we derived for the isolated galaxy systems traces 
 the local halo of the isolated primary galaxies, 
 while the $\sigma_{sys}$ for galaxy clusters traces the extended cluster halo, much larger than the BCG halo.
This separation is consistent with the interpretation of cluster simulations. 
\citet{dol10} demonstrate that there are two dynamically well-distinct stellar components 
 in simulated galaxy clusters: a component with a small velocity dispersion traces the local halo of the BCGs (or cD galaxies) and 
 the other component represents a diffuse stellar population governed by the cluster halo. 
In this point of view, the $\sigma_{sys}$ of galaxy clusters may be consistent with the $\sigma$ of the diffuse stellar population, and the $\sigma_{sys}$ of isolated galaxy systems may be consistent with the $\sigma$ of the primary galaxy halos. 

To test this hypothesis further, 
 observations of other tracers for the primary galaxies of clusters and isolated galaxies are required. 
Strong lensing observations provide excellent constraints on the mass associated with the galaxy and cluster halo, respectively (e.g., \citealp{mon17}). 
For local BCGs or primary galaxies, 
 observations for other dynamical mass tracers including globular clusters and planetary nebulae 
 enable us to constraint the mass within the local halo of the primary galaxies (e.g., \citealp{ko17,lon18a,lon18b}). 
 
\section{Summary and Conclusion}

We construct a complete sample of isolated galaxies hosting faint satellites ($\Delta r > 2$ mag)
 by applying the FoF algorithm to the SDSS DR12 spectroscopic galaxy catalog. 
We first divide the sample according to the color of the primary galaxies:
 red primary galaxies with $(g-r)_{0} > 0.85$ and blue primary galaxies with $(g-r)_{0} \leq 0.85$.
Based on the large sample, 
 we stack the galaxy systems depending on the physical properties (i.e., $M_{r}, M_{*}, \sigma_{*, prim}$) 
 of their primary galaxies to derive the system velocity dispersion. 
Then, we investigated the relation between the system velocity dispersion 
 and the physical properties of the primary galaxies. 
The main results are summarized as follows. 

\begin{itemize}

\item Velocity dispersions of the galaxy systems ($\sigma_{sys}$) show a strong correlation 
 with the central stellar velocity dispersion of the primary galaxies ($\sigma_{*, prim})$. 
In particular, in the case of systems with the red primary galaxies, 
 $\sigma_{sys}$ is directly proportional to $\sigma_{*, prim}$ with a slope of 1. 
The $\sigma_{sys}$ for the blue primary galaxies is also correlated with $\sigma_{*,prim}$, 
 but with a shallower slope  ($\alpha = 0.74 \pm 0.09$). 
 
\item $\sigma_{sys}$ is also proportional to the luminosity and stellar mass of the primary galaxies. 
In general, the more massive systems host red primary galaxies. 

\item Because there is a power-law relation between $\sigma_{sys}$ and the dark matter halo mass ($M_{DM}$), the physical properties of primary galaxies are correlated with their dark matter halo masses. We derive the relations between $\sigma_{*, prim}$ and $M_{DM}$ (as well as $M_{*}$) 
 for the further comparison with various models. 
 
\item We compare the $\sigma_{sys} - \sigma_{*,prim}$ relation of the isolated galaxy systems 
 with the same relation for galaxy clusters from \citet{soh20}. 
The relation for the galaxy cluster shows a much steeper slope.
The different slopes of the relations suggest that 
 the $\sigma_{sys}$ for our target systems and the clusters trace different halos, 
 i.e., the local halo of the primary galaxies and the halo of the entire clusters, respectively. 
 
\end{itemize}

In conclusion, 
 the stellar velocity dispersion of a galaxy is an efficient and robust tracer 
 for its dark matter halo mass. 
We highlight that the stellar velocity dispersion as a robust spectroscopic measure 
 which can be measured from the future large spectroscopic surveys like DESI, 4MOST, and Subaru/PFS.  
In the near future, 
 wide-field imaging and spectroscopic observations will enable the exploration of 
 dark matter halo properties based on statistical analysis of lensing observations. 
Combining this dark matter mass estimates with 
 a large sample of stellar velocity dispersion measurements 
 would be an important test for galaxy and structure formation models.

\acknowledgments
This work was supported by the National Research Foundation grant funded by the Korean Government (NRF-2019R1A2C2084019). 
We thank an anonymous referee for useful comments and Margaret Geller for her insightful advice on this project. 
We thank Brian S. Cho for his help in improving the English in the manuscript. 
We also acknowledge Antonaldo Diaferio and Ken Rines for helpful discussions. 
J.S. is supported by the CfA Fellowship.  

This research has made use of NASA's Astrophysics Data System Bibliographic Services.
Funding for the Sloan Digital Sky Survey IV has been provided by the Alfred P. Sloan Foundation, the US Department of Energy Office of Science, and the Participating Institutions. SDSS-IV acknowledges
support and resources from the Center for High-Performance Computing at
the University of Utah. The SDSS website is www.sdss.org.

SDSS-IV is managed by the Astrophysical Research Consortium for the 
Participating Institutions of the SDSS Collaboration including the 
Brazilian Participation Group, the Carnegie Institution for Science, 
Carnegie Mellon University, the Chilean Participation Group, the French Participation Group, Harvard-Smithsonian Center for Astrophysics, 
Instituto de Astrof\'isica de Canarias, The Johns Hopkins University, Kavli Institute for the Physics and Mathematics of the Universe (IPMU)/University of Tokyo, the Korean Participation Group, Lawrence Berkeley National Laboratory, 
Leibniz Institut f\"ur Astrophysik Potsdam (AIP),  
Max-Planck-Institut f\"ur Astronomie (MPIA Heidelberg), 
Max-Planck-Institut f\"ur Astrophysik (MPA Garching), 
Max-Planck-Institut f\"ur Extraterrestrische Physik (MPE), 
National Astronomical Observatories of China, New Mexico State University, 
New York University, University of Notre Dame, 
Observat\'ario Nacional / MCTI, The Ohio State University, 
Pennsylvania State University, Shanghai Astronomical Observatory, 
United Kingdom Participation Group,
Universidad Nacional Aut\'onoma de M\'exico, University of Arizona, 
University of Colorado Boulder, University of Oxford, University of Portsmouth, 
University of Utah, University of Virginia, University of Washington, University of Wisconsin, 
Vanderbilt University, and Yale University.

\end{document}